# Quadrature Compressive Sampling for Radar Signals

Feng Xi, Shengyao Chen, and Zhong Liu

***Abstract*—Quadrature sampling has been widely applied in coherent radar systems to extract in-phase and quadrature (*I* and *Q*) components in the received radar signal. However, the sampling is inefficient because the received signal contains only a small number of significant target signals. This paper incorporates the compressive sampling (CS) theory into the design of the quadrature sampling system, and develops a quadrature compressive sampling (QuadCS) system to acquire the *I* and *Q* components with low sampling rate. The QuadCS system first randomly projects the received signal into a compressive bandpass signal and then utilizes the quadrature sampling to output compressive *I* and *Q* components. The compressive outputs are used to reconstruct the *I* and *Q* components. To understand the system performance, we establish the frequency domain representation of the QuadCS system. With the waveform-matched dictionary, we prove that the QuadCS system satisfies the restricted isometry property with overwhelming probability. For $K$ target signals in the observation interval $T$, simulations show that the QuadCS requires just $\mathcal{O}\left(K\log\left(BT/K\right)\right)$ samples to stably reconstruct the signal, where $B$ is the signal bandwidth. The reconstructed signal-to-noise ratio decreases by 3dB for every octave increase in the target number $K$ and increases by 3dB for every octave increase in the compressive bandwidth. Theoretical analyses and simulations verify that the proposed QuadCS is a valid system to acquire the *I* and *Q* components in the received radar signals.**

***Index Terms*—Analog-to-digital conversion, Compressive sampling, Quadrature sampling, Restricted isometry property, Sparse signal reconstruction.**

## I. Introduction

IN current radar systems, coherent detectors are often formulated from two parallel baseband channels, *in-phase* and *quadrature* (denoted as *I* and *Q*) components, of radar RF echo signals. Quadrature sampling is widely used for digital *I* and *Q* component acquisition because it overcomes the inherent difficulty in obtaining perfect *I* and *Q* channel balance of analog demodulator [1~2]. For the bandpass signal with center frequency $f_c$ and bandwidth $B$, the quadrature sampling theorem [1] states that the digital *I* and *Q* components can be acquired with sampling frequency

$$f_{IF} = \frac{4f_L + 2B}{4l+1} \tag{1}$$

where $l$ is a positive integer satisfying $l \le \lfloor f_L/2B \rfloor$ and $f_L = f_c - B/2$. With appropriate setting of $f_c$, the minimum sampling rate $2B$ can be allocated. The requirement (1) has become a serious bottleneck in the development of wideband/ ultrawideband radar systems.

The newly introduced compressive sampling (CS) [3~5], or compressive sensing, brings us a new concept on the low-rate data acquisition. The CS theory exploits the sparsity of signals and samples signals closer to their information rate instead of their bandwidth. With high probability, the CS can recover sparse signals from far fewer samples or measurements than the Nyquist samples. The fewer samples lead to a reduced sampling rate and, hence, to a reduced use of analog-to-digital converter (ADC) resources. Along with the CS theory, several schemes have been proposed to implement the CS or analog-to-information conversion (A2I) of the analog signals, including random sampling[6], random filtering[7], random demodulation (RD) [8,9], segmented compressed sampling [10], Xampling[11], and so on. These A2I schemes are generic and applicable to the signals sparse in frequency, time or time-frequency domains. In some applications, for example, in the sampling of radar signals discussed in this paper, they may not be able directly to obtain the discrete baseband components of the RF signals. In addition, we may know in advance some information of transmitting waveforms, for example, radio frequency, baseband waveform, modulation mode and so on. Incorporating the *a priori* information into analog CS design may reduce A2I's implementation complexity and thus improve A2I's performance.

In this paper, we merge the A2I technique [8, 9] into the quadrature sampling and develop a quadrature compressive sampling (QuadCS) system for discrete *I* and *Q* acquisition of RF signals. The *a priori* information of radar transmitting signal, radar frequency and waveform, is incorporated into the QuadCS design. As shown in Fig.1, the QuadCS consists of a low-rate sampling subsystem, a quadrature demodulation subsystem and an *I* and *Q* component reconstruction subsystem. The first subsystem randomly projects the received RF or intermediate-frequency signal to a compressive bandpass signal with bandwidth $B_{cs}$ and samples the compressive signal with the low-rate ADC. Because the bandwidth $B_{cs}$ can be set to be much smaller than that of the received signal, the sampling rate is thus greatly reduced. The second subsystem performs the

Manuscript received April 23, 2013. This work was supported in part by the National Science Foundation of China under Grant 61171166 and 61101193. The source of this paper was partly presented at the 2011 International Conference on Wireless Communications and Signal Processing, Nov.2011.

Feng Xi, Shengyao Chen and Zhong Liu are with the Department of Electrical Engineering, Nanjing University of Science and Technology, Nanjing, 210094, China. (e-mail: xifeng.njust@gmail.com, chen_shengyao @163.com, and eezliu@njust.edu.cn).

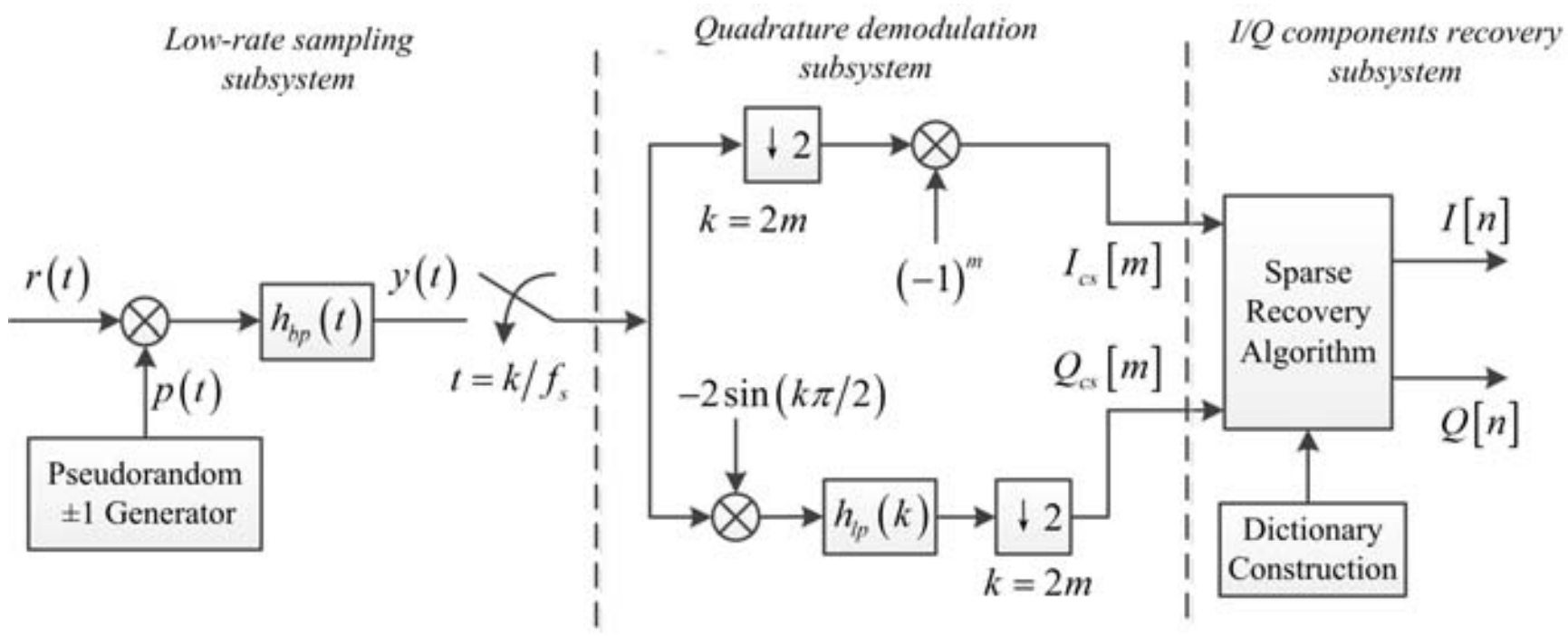

Fig.1 The structure of the proposed QuadCS system.

same as the traditional digital quadrature demodulation to extract the compressive *I* and *Q* components from the output of the low-rate sampling subsystem. With the compressive *I* and *Q* measurements provided by the quadrature demodulation, the last subsystem reconstructs the baseband *I* and *Q* components by solving a $l_1$-norm optimization problem [12]. For radar signals, the waveform-matched dictionary [13] is considered, which consists of different time-shifted versions of the transmitting signal at the Nyquist sampling grids of the observation interval. Different from other A2I schemes, the QuadCS works on bandpass signals and can directly derive the baseband *I* and *Q* components for subsequent processing.

Other contributions of this paper are the following. i) We derive the frequency-domain representation of the QuadCS system for its valid analysis. The frequency-domain representation discloses the inherent relations among radar signals, the QuadCS operations and the compressive measurements. ii) Based on the frequency-domain representation, we prove that the restricted isometry property (RIP) [3,14], *i.e.*, the sufficient condition for sparse signal recovery, can be satisfied for proper setting of QuadCS system parameters. It is found that $B_{cs} = \mathcal{O}\left(\left(K^2/T\right)\log\left(BT/K\right)\right)$ is sufficient for high-probability recovery of the $K$ target signals in the observation interval $T$, where $B$ is the signal bandwidth. iii) We conduct extensive simulations on performance of the QuadCS system in noise/noise-free cases, with two types of radar signals, linear-frequency modulated (LFM) signals and phase-coded signals. An empirical formula $B_{cs} \geq 1.78\left(K/T\right)\log\left(BT/K\right)$ is formulated, which guarantees the reconstruction of the radar baseband signals with high probability. For the signals contaminated in Gaussian noise, the simulation results show that the reconstructed signal-to-noise ratio (SNR) decreases by 3dB for every octave increase in the target number $K$. The noise-folding phenomenon [15,16] also exists in the QuadCS system, *i.e.* the reconstructed SNR increases by 3dB for every octave increase in the compressive bandwidth. When the time-delays of the real targets are not at the Nyquist-sampling grids, the real target signals deviate from the assumed signal model and the reconstruction performance degenerates. This problem is called as the *basis mismatch* in CS [17]. However, the simulations show that the mismatch has little effect on the estimation of the target positions from the point of view of the time-delay estimation. Our theoretical and simulated study shows that the QuadCS system is an effective sampling system for low-rate acquisition of sparse radar signals.

The rest of this paper is organized as follows. Signal model and sparse representation are given in Section II. The QuadCS system and its operation principle are explained in Section III. The frequency-domain representation of the QuadCS system is derived in Section IV. We prove in Section V that the measurement matrix resulting from the proposed QuadCS satisfies the restricted isometry property and, therefore, the QuadCS is an effective A2I system for radar signals. Section VI shows our simulation results and Section VII concludes the paper.

Before proceeding, we briefly introduce some mathematical notation that we will use throughout the paper. To avoid confusion, a real signal, a complex signal and the Fourier transform of a signal are represented as $s(t)$, $\tilde{s}(t)$ and $\hat{S}(j\Omega)$, respectively. Bold letters denote the matrices or vectors. $\mathrm{Re}\{\cdot\}$ and $\mathrm{Im}\{\cdot\}$ take the real part and imaginary part of $\{\cdot\}$. $(\cdot)^*$, $(\cdot)^{\mathrm{T}}$ and $(\cdot)^{\mathrm{H}}$ represent complex conjugation, transposition, and conjugate transposition of $(\cdot)$, respectively. $\|\cdot\|_2$ denotes the Euclidean norm, and $\|\cdot\|_1$ is $l_1$-norm.

## II. Signal Model and Sparse Representation

### A. Signal Model

In radar systems, the received radio signals are often firstly down-converted to the IF signals for the subsequent processing. For $K$ non-fluctuating point targets, the received echo signal at the IF $f_0$ can be represented as

$$r(t) = \sum_{k=1}^{K} \nu_k a(t-t_k)\cos\left[2\pi f_0(t-t_k) + \phi(t-t_k) + \varphi_k\right] \quad (2)$$

where $a(t)$ and $\phi(t)$ are a priori known envelope and phase of the radar transmitting waveform, respectively; $t_k$, $\nu_k$ and $\varphi_k$ are the time delay, gain coefficient and phase offset of the $k$-th target, respectively. The signal (2) has a bandpass spectrum with center frequency $f_0$ and bandwidth $B$, where $B$ is the bandwidth of transmitting waveform and $f_0 > B/2$.

Denote $\tilde{s}_0(t) = a(t)e^{j\phi(t)}$ as the complex baseband signal of the radar waveform and $\tilde{\nu}_k = \nu_k e^{j\varphi_k'}$ with $\varphi_k' = \varphi_k - 2\pi f_0 t_k$. The complex envelope $\tilde{s}(t)$ of the received signal $r(t)$ is given by

$$\tilde{s}(t) = \sum_{k=1}^{K} \tilde{\nu}_k \tilde{s}_0(t-t_k) \quad (3)$$

Then the baseband *I* and *Q* components of the received signal $r(t)$ are defined as

$$I(t) = \mathrm{Re}\{\tilde{s}(t)\} = \sum_{k=1}^{K} \nu_k a(t-t_k)\cos\left[\phi(t-t_k) + \varphi_k'\right] \quad (4)$$

$$Q(t) = \mathrm{Im}\{\tilde{s}(t)\} = \sum_{k=1}^{K} \nu_k a(t-t_k)\sin\left[\phi(t-t_k) + \varphi_k'\right] \quad (5)$$

The target information is completely included in the complex envelope $\tilde{s}(t)$ or the *I* and *Q* components, $I(t)$ and $Q(t)$. The aim of the digital quadrature demodulation is to acquire the digital *I* and *Q* components from the IF signal $r(t)$. In the observation interval $T$, we can obtain at least $2BT$ samples of $I(t)$ and $Q(t)$ or $BT$ complex samples of $\tilde{s}(t)$ by the digital quadrature demodulation.

### B. Waveform-Matched Dictionary

To obtain the samples of $\tilde{s}(t)$ by the CS theory, $\tilde{s}(t)$ should be sparse in some dictionary. In radar applications, the transmitting waveforms are known in advance. A natural one is the waveform-matched dictionary [13]. For the radar with the baseband signal $\tilde{s}_0(t)$ of the bandwidth $B$, let $\tau_0$ be its Nyquist sampling interval, $\tau_0 = 1/B$, and let $N$ be the Nyquist sampling number of the receiver signals in the observation interval $T$, *i.e.*, $N = \lfloor T/\tau_0 \rfloor = \lfloor BT \rfloor$. Then the waveform-matched dictionary consists of all the time-shifted versions of $\tilde{s}_0(t)$ at the Nyquist-sampling grids $\{\tau_0, \cdots, N\tau_0\}$, *i.e.*, $\{\tilde{\psi}_n(t) | \tilde{\psi}_n(t) = \tilde{s}_0(t - n\tau_0), n = 1, \cdots, N\}$. In the waveform-matched dictionary, the time- delay axis is discretized with the resolution $\tau_0$. The discretization is reasonable, because the time resolution of the bandlimited baseband signal $\tilde{s}_0(t)$ is limited to $1/B$.

Assume that the target delays are at the integral multiples of $\tau_0$, *i.e.*, $t_k \in \{\tau_0, \cdots, N\tau_0\}$. Given the waveform-matched dictionary, the complex envelope $\tilde{s}(t)$ in (3) can be represented as

$$\begin{aligned}\tilde{s}(t) &= \sum_{n=1}^{N} \tilde{v}_n \tilde{\psi}_n(t) \\ &= \tilde{\boldsymbol{\Psi}}(t)\tilde{\mathbf{v}}\end{aligned} \tag{6}$$

where $\tilde{\mathbf{v}} = [\tilde{v}_1, \tilde{v}_2, \cdots, \tilde{v}_N]^{\mathrm{T}}$ and $\tilde{\boldsymbol{\Psi}}(t) = [\tilde{\psi}_1(t), \tilde{\psi}_2(t), \cdots, \tilde{\psi}_N(t)]$ are called the coefficient vector and the dictionary vector, respectively. For $K$ targets, there are $N-K$ zero coefficients. When $K \ll N$, $\tilde{s}(t)$ is said to be $K$-sparse in the waveform-matched dictionary. The sparsity level $K$ exactly equals the number of the targets.

With the sparse representation in the waveform-matched dictionary, this paper focuses on the design of a compressive sampling system, QuadCS system, to acquire the compressive *I* and *Q* components from the IF waveform $r(t)$ and recover the Nyquist sampling *I* and *Q* components. With the compressive sampling scheme, the samples of $I(t)$ and $Q(t)$ will be much less than $2BT$ by the digital quadrature demodulation.

## III. Quadrature Compressive Sampling (QuadCS) System

In this section, we examine the QuadCS structure shown in Fig.1 and analyze its operating principle.

### A. Low-Rate Sampling Subsystem

The low rate sampling subsystem is similar to the random demodulation scheme in [8, 9] and implements the low-rate sampling of the input bandpass analog signals. Since the signal in processing is a bandpass one with the known IF frequency rather than a lowpass signal as that in the random demodulation, the component operations of the low rate sampling subsystem are different as explained in the following.

The received IF signal $r(t)$ is firstly mixed by a random-binary signal $p(t)$

$$p(t) = \varepsilon_k, \quad t \in [k/B_p, (k+1)/B_p), k = 0, 1, \cdots \tag{7}$$

where $B_p \geq B$ and $\varepsilon_k = 1$ or $-1$. The $p(t)$ is called as chipping sequence [9], which alternates between +1 and -1 randomly at or above the Nyquist rate $1/\tau_0$ of the baseband signal. The mixing operation will spread the frequency content of the baseband signal to full spectrum of $p(t)$. Different from that in random demodulation, the chipping rate is determined by the passband width of the received signal $r(t)$ instead of the upper frequency of $r(t)$.

The analog bandpass filter $h_{bp}(t)$ is centered at the frequency $f_0$ with bandwidth $B_{cs} \ll B$. The output $y(t)$ of the filter is a compressive bandpass signal with center frequency $f_0$ and the bandwidth $B_{cs}$

$$\begin{aligned}y(t) &= h_{bp}(t) * (p(t)r(t)) \\ &= \int_{-\infty}^{\infty} h_{bp}(\tau) p(t-\tau) r(t-\tau) d\tau \\ &= \mathrm{Re}\left\{\tilde{s}_{cs}(t) e^{j2\pi f_0 t}\right\}\end{aligned} \tag{8}$$

where $\tilde{s}_{cs}(t)$ is the compressive complex envelope

$$\tilde{s}_{cs}(t) = \int_{-\infty}^{+\infty} h_{bp}(\tau) e^{-j2\pi f_0 \tau} p(t-\tau) \tilde{s}(t-\tau) d\tau \tag{9}$$

with $I_{cs}(t) = \mathrm{Re}\{\tilde{s}_{cs}(t)\}$ and $Q_{cs}(t) = \mathrm{Im}\{\tilde{s}_{cs}(t)\}$ denoting the compressive *I* and *Q* components, respectively.

The compressive bandpass signal $y(t)$ is sampled with a low-rate ADC according to the bandpass sampling theorem [1]. Let the lower and upper band edge of $y(t)$ be $f_L = f_0 - B_{cs}/2$ and $f_H = f_0 + B_{cs}/2$, the sampling frequency $f_{IF}^{cs}$ can be chosen as

$$f_{IF}^{cs} = (4f_L + 2B_{cs})/(4l+1) \tag{10}$$

with $l$ be a positive integer satisfying $l \leq \lfloor f_L/2B_{cs} \rfloor$. The minimum sampling frequency $f_{IF}^{cs}$ is $2B_{cs}$, provided that $f_L/2B_{cs}$ is an integer, or equivalently, $f_0 = (2l + 1/2)B_{cs}$ ($l = 1, 2, \cdots$). The output of the low-rate sampling subsystem is a sequence of the discrete-time samples $\{y[k], k = 0, 1, 2, \cdots\}$ with

$$\begin{aligned}y[k] &= y(k/f_{IF}^{cs}) \\ &= \begin{cases}(-1)^{k/2} I_{cs}(k/f_{IF}^{cs}) & k \text{ is even} \\ (-1)^{(k+1)/2} Q_{cs}(k/f_{IF}^{cs}) & k \text{ is odd}\end{cases}\end{aligned} \tag{11}$$

The proposed sampling frequency $f_{IF}^{cs}$ is much lower than that of directly sampling (1) because $B_{cs}$ can be set to be much smaller than $B$. However, to guarantee the successful recovery of the complex envelope $\tilde{s}(t)$ from the low-rate samples, we have to properly choose the bandwidth $B_{cs}$ of the bandpass filter. As discussed in Section V, the selection will depend on the bandwidth $B$, the observation interval $T$ and the sparse level $K$.

### B. Quadrature Demodulation Subsystem

The quadrature demodulation subsystem is used to extract digital compressive *I* and *Q* components from the low-rate sampling sequence $y[k]$. Its operation is the same as in traditional quadrature sampling [2]. The digital lowpass filter $h_{lp}$ in Fig.1 plays in role of aligning the *Q* component with *I* component at the same sampling instants. The digital compressive *I* component is obtained by firstly down-sampling $y[k]$ by a factor of 2 and then multiplying the down-sampling sequence $(-1)^m I_{cs}\left(2m/f_{IF}^{cs}\right)$ with the sequence $(-1)^m$

$$I_{cs}[m] = I_{cs}(mT_{cs}) \qquad m = 0,1,2,\cdots \tag{12}$$

where $T_{cs} = 2/f_{IF}^{cs}$ . The digital compressive $Q$ component is obtained by firstly digital demodulation of $y[k]$ through $-2\sin(k\pi/2)$, then lowpass filtering the demodulation output and finally down-sampling the filtering output by a factor of 2

$$Q_{cs}[m] = Q_{cs}(mT_{cs}) \qquad m = 0,1,2,\cdots \tag{13}$$

With this structure, we get the compressive samples of *I* and *Q* components, $I_{cs}[m]$ and $Q_{cs}[m]$, or equivalently the complex samples $\tilde{s}_{cs}[m] = \tilde{s}_{cs}(mT_{cs})$ , $\tilde{s}_{cs}[m] = I_{cs}[m] + jQ_{cs}[m]$ , of the compressive complex envelope $\tilde{s}_{cs}(t)$.

### C. Reconstruction Subsystem of I and Q Components

With the sparse representation (6), we can express the compressive complex envelope $\tilde{s}_{cs}(t)$ (9) as

$$\tilde{s}_{cs}(t) = \sum_{n=1}^{N} \tilde{\nu}_n \int_{-\infty}^{+\infty} h_{bp}(\tau) e^{-j2\pi f_0 \tau} p(t-\tau) \tilde{\psi}_n(t-\tau) d\tau \tag{14}$$

Then the complex samples $\tilde{s}_{cs}[m]$ will be

$$\tilde{s}_{cs}[m] = \sum_{n=1}^{N} \tilde{\nu}_n \int_{-\infty}^{+\infty} h_{bp}(\tau) e^{-j2\pi f_0 \tau} p(mT_{cs}-\tau) \tilde{\psi}_n(mT_{cs}-\tau) d\tau \tag{15}$$

In the observation interval $T$ , we obtain $M = \lfloor T/T_{cs} \rfloor$ complex samples of $\tilde{s}_{cs}(t)$. Since $B_{cs} \ll B$ , the number of the complex samples is much less than $BT$ . Define the measurement vector $\tilde{\mathbf{s}}_{cs}$ and the measurement matrix $\tilde{\mathbf{M}}$ in time domain, respectively, as

$$\tilde{\mathbf{s}}_{cs} = \left[\tilde{s}_{cs}[0], \cdots, \tilde{s}_{cs}[M-1]\right]^{\mathrm{T}} \tag{16}$$

and

$$\tilde{\mathbf{M}} = \left[\tilde{M}_{mn}\right] \in \mathbb{C}^{M\times N} \tag{17}$$

where $\tilde{M}_{mn} = \int_{-\infty}^{+\infty} h_{bp}(\tau) e^{-j2\pi f_0 \tau} p(mT_{cs}-\tau) \tilde{\psi}_n(mT_{cs}-\tau) d\tau$. We can express (15) in the matrix form

$$\tilde{\mathbf{s}}_{cs} = \tilde{\mathbf{M}}\tilde{\mathbf{v}} \tag{18}$$

The QuadCS system is completely described by the matrix $\tilde{\mathbf{M}}$ . With the sparse representation, the recovery of the complex envelope $\tilde{s}(t)$ is equivalently to reconstruct the sparse coefficient vector $\tilde{\mathbf{v}}$ . If $M \geq N$ , $\tilde{\mathbf{v}}$ can be found by directly solving (18). In the observation interval $T$ , $M < N$ and then directly solving (18) is an ill-posed problem. However, the CS theory states that if the matrix $\tilde{\mathbf{M}}$ satisfies the RIP [14], it is indeed possible to reconstruct the $K$ -sparse $\tilde{\mathbf{v}}$ . The matrix $\tilde{\mathbf{M}}$ is said to satisfy the RIP with parameter $\delta_K$ if the following inequality holds for every $K$ -sparse $\tilde{\mathbf{v}} \in \mathbb{C}^N$

$$(1-\delta_K)\|\tilde{\mathbf{v}}\|_2^2 \leq \left\|\tilde{\mathbf{M}}\tilde{\mathbf{v}}\right\|_2^2 \leq (1+\delta_K)\|\tilde{\mathbf{v}}\|_2^2 \tag{19}$$

The RIP states that the measurement matrix approximately preserves the norm of any $K$ -sparse vector living in $\mathbb{C}^N$ when mapping it to a $M$ -dimensional space ( $M < N$ ). Thus, the information of the $K$ -sparse signal is preserved, making the reconstruction possible.

For the matrix $\tilde{\mathbf{M}}$ with the RIP, the sparse coefficient vector $\tilde{\mathbf{v}}$ can be recovered by solving the following $l_1$ -norm optimization problem

$$\begin{cases} \min \|\tilde{\mathbf{v}}\|_1 \\ \text{s.t. } \tilde{\mathbf{s}}_{cs} = \tilde{\mathbf{M}}\tilde{\mathbf{v}} \end{cases} \tag{20}$$

The above three subsections describe the compressive sampling and reconstruction of the received radar echo waveform in noise-free case. In practice, the received echoes are contaminated in noise. In this case, the measurement vector by the QuadCS system is given by

$$\tilde{\mathbf{s}}_{cs} = \tilde{\mathbf{M}}\tilde{\mathbf{v}} + \tilde{\mathbf{n}}_{cs} \tag{21}$$

where $\tilde{\mathbf{n}}_{cs}$ denotes the compressive samples of the additive noise in $r(t)$ . The representation is reasonable because the QuadCS is a linear system. The reconstruction of the sparse coefficient vector $\tilde{\mathbf{v}}$ in noise case is to solve

$$\begin{cases} \min \|\tilde{\mathbf{v}}\|_1 \\ \text{s.t. } \left\|\tilde{\mathbf{s}}_{cs} - \tilde{\mathbf{M}}\tilde{\mathbf{v}}\right\|_2 \leq \varepsilon \end{cases} \tag{22}$$

where $\varepsilon$ is a small parameter determined by the noise distribution.

There are a wide variety of approaches to solve (20) and (22), including the greedy iteration algorithms [18,19] and convex optimization algorithms [20,21] (see [22] for a review). In the simulation study, we use the basis pursuit (BP) algorithm and the basis pursuit denoise (BPDN) algorithm [23, 24] in the SGPL1 toolbox [25] to find the sparse vector $\tilde{\mathbf{v}}$ .

## IV. Frequency-Domain Representation of the QuadCS System

To study the performance of the QuadCS system, we derive in this section its frequency-domain representation, which establishes in matrix form the explicit relations among the baseband complex envelope, the system operations and the compressive measurements.

To simplify the analysis, we make the following assumptions:

a) The bandpass filter $h_{bp}(t)$ is ideal and its frequency response is given by $\hat{H}_{bp}(j\Omega) = \hat{H}_{lp}\left(j(\Omega - 2\pi f_0)\right) + \hat{H}_{lp}\left(j(\Omega + 2\pi f_0)\right)$ , where $\hat{H}_{lp}(j\Omega)$ is the frequency response of an ideal lowpass filter

$$\hat{H}_{lp}(j\Omega) = \begin{cases} B/B_{cs} & -\pi B_{cs} \leq \Omega \leq \pi B_{cs} \\ 0 & else \end{cases} \tag{23}$$

The gain $B/B_{cs}$ is to keep the signal energy unchanged after filtering.

b) The spectral-spreading signal $p(t)$ is periodic with the observation interval $T$ as its period and its alternating rate $B_p$ is set to be $B_p = B$ .

c) The bandpass sampling frequency for the compressive bandpass signal $y(t)$ is set as the minimum sampling frequency, $f_{IF}^{cs} = 2B_{cs}$ . In this case, $T_{cs} = 1/B_{cs}$ .

d) The observation interval $T$ satisfies $T = N/B = M/B_{cs}$ , which ensures that the sample numbers of the observed signal by the Nyquist sampling and by the QuadCS are of integer. Furthermore, $N - M$ is assumed to be even.

Firstly, we consider the discrete-time Fourier transform (DTFT) $\hat{s}_{cs}(e^{j\omega})$ of the complex sampling sequence $\tilde{s}_{cs}[m] = \tilde{s}_{cs}(mT_{cs})$. Denote $\hat{S}_{cs}(j\Omega)$ as the Fourier transform of the compressive complex envelope $\tilde{s}_{cs}(t)$ . Define $\tilde{s}_n^p(t) = p(t)\tilde{\psi}_n(t)$ and $\hat{S}_n^p(j\Omega) = \int_{-\infty}^{+\infty} p(t)\tilde{\psi}_n(t) e^{-j\Omega t} dt$ as its Fourier transform. By the convolution theorem and (23), we have the Fourier transform $\hat{S}_{cs}(j\Omega)$ as

$$\hat{S}_{cs}(j\Omega)=\int_{-\infty}^{+\infty}s_{cs}(t)e^{-j\Omega t}dt$$
$$=\sum_{n=1}^{N}\tilde{\nu}_n\hat{H}_{lp}(j\Omega)\hat{S}_n^p(j\Omega) \tag{24}$$
$$=\begin{cases}(B/B_{cs})\sum_{n=1}^{N}\tilde{\nu}_n\hat{S}_n^p(j\Omega) & \Omega\in[-\pi B_{cs},\pi B_{cs}]\\ 0 & else\end{cases}$$

Then $\hat{s}_{cs}(e^{j\omega})$ is a linear combination of $1/T_{cs}$ -shifted copies of $\hat{S}_{cs}(j\Omega)$

$$\hat{s}_{cs}(e^{j\omega})=\sum_{m=-\infty}^{+\infty}\tilde{s}_{cs}[m]e^{-j\omega m}$$
$$=\frac{1}{T_{cs}}\sum_{k=-\infty}^{+\infty}\hat{S}_{cs}\left(j\left(\frac{\omega}{T_{cs}}-\frac{2\pi k}{T_{cs}}\right)\right) \tag{25}$$
$$=B\sum_{n=1}^{N}\tilde{\nu}_n\sum_{k=-\infty}^{+\infty}\hat{S}_n^p\left(j\left(\frac{\omega}{T_{cs}}-\frac{2\pi k}{T_{cs}}\right)\right)$$

Similarly, letting $\tilde{s}_n^p[l]$ be the samples of $\tilde{s}_n^p(t)$ at the Nyquist sampling rate $T_{nyq}=1/B$, we have the DTFT $\hat{s}_n^p(e^{j\omega})$ of $\tilde{s}_n^p[l]$ as

$$\hat{s}_n^p(e^{j\omega})=\sum_{l=-\infty}^{+\infty}\tilde{s}_n^p[l]e^{-j\omega l}$$
$$=\frac{1}{T_{nyq}}\sum_{k=-\infty}^{+\infty}\hat{S}_n^p\left(j\left(\frac{\omega}{T_{nyq}}-\frac{2\pi k}{T_{nyq}}\right)\right) \tag{26}$$

Combining (26) with (25), we have $\hat{s}_{cs}(e^{j\omega})$ in one period as

$$\hat{s}_{cs}(e^{j\omega})=\sum_{n=1}^{N}\tilde{\nu}_n\hat{s}_n^p\left(e^{jM\omega/N}\right)\qquad(\omega\in(-\pi,\pi]) \tag{27}$$

(27) is the equivalent frequency-domain representation of (14).

Secondly, we perform discrete measurements of $\tilde{s}_{cs}(t)$ in frequency domain by sampling (27). Let $\omega_m=-\pi+2\pi m/M$ ( $m=0,1,\cdots,M-1$ ). Define the measurement vector $\hat{\mathbf{s}}_{cs}$ and the measurement matrix $\hat{\mathbf{M}}$ in frequency domain, respectively, as

$$\hat{\mathbf{s}}_{cs}=M^{-1/2}\left[\hat{s}_{cs}(e^{j\omega_0}),\hat{s}_{cs}(e^{j\omega_1}),\ldots,\hat{s}_{cs}(e^{j\omega_{M-1}})\right]^{\mathrm{T}}\in\mathbb{C}^M \tag{28}$$

and

$$\hat{\mathbf{M}}=M^{-1/2}\begin{bmatrix}\hat{s}_1^p\left(e^{jM\omega_0/N}\right) & \cdots & \hat{s}_N^p\left(e^{jM\omega_0/N}\right)\\ \vdots & & \vdots\\ \hat{s}_1^p\left(e^{jM\omega_{M-1}/N}\right) & \cdots & \hat{s}_N^p\left(e^{jM\omega_{M-1}/N}\right)\end{bmatrix} \tag{29}$$

Then we have

$$\hat{\mathbf{s}}_{cs}=\hat{\mathbf{M}}\tilde{\mathbf{v}} \tag{30}$$

(30) gives the frequency domain measurements of the QuadCS system, corresponding to that in (18).

Finally, we conduct the factorization of the measurement matrix $\hat{\mathbf{M}}$, which is helpful to derive the RIP conditions in next section. Note that $\tilde{s}_n^p[l]=p[l]\tilde{\psi}_n[l]$ and $p[l]$ is a period- $N$ sequence at the Nyquist sampling rate. Then $p[l]$ has its discrete Fourier series (DFS) expansion as

$$p[l]=N^{-1}\sum_{k=0}^{N-1}\hat{c}_p[k]e^{j2\pi kl/N} \tag{31}$$

where $\hat{c}_p[k]=\sum_{l=0}^{N-1}p[l]e^{-j2\pi kl/N}$ . Then $\hat{s}_n^p(e^{j\omega})$ can also be represented as

$$\hat{s}_n^p(e^{j\omega})=N^{-1}\sum_{l=-\infty}^{+\infty}\left(\sum_{k=0}^{N-1}\hat{c}_p[k]e^{j2\pi kl/N}\right)\tilde{\psi}_n[l]e^{-j\omega l}$$
$$=N^{-1}\sum_{k=0}^{N-1}\hat{c}_p[k]\sum_{l=-\infty}^{+\infty}\tilde{\psi}_n[l]e^{-j(\omega-2\pi k/N)l} \tag{32}$$
$$=N^{-1}\sum_{k=0}^{N-1}\hat{c}_p[k]\hat{\psi}_n\left(e^{j(\omega-2\pi k/N)}\right)$$

where $\hat{\psi}_n(e^{j\omega})$ is the DTFT of the Nyquist sampling sequence $\tilde{\psi}_n[l]$. The elements $\hat{M}_{mn}$ of $\hat{\mathbf{M}}$ will be

$$\hat{M}_{mn}=M^{-1/2}\hat{s}_n^p(e^{j\omega})\Big|_{\omega=M\omega_m/N}$$
$$=M^{-1/2}N^{-1}\sum_{k=0}^{N-1}\hat{c}_p[k]\hat{\psi}_n\left(e^{j(M\omega_m/N-2\pi k/N)}\right) \tag{33}$$
$$=M^{-1/2}N^{-1}\sum_{l=0}^{N-1}\hat{c}_p\left[(N-M)/2+m-l\right]\hat{\psi}_n\left(e^{j(-\pi+2\pi l/N)}\right)$$

Define the matrices $\mathbf{R}=[r_{ml}]\in\mathbb{R}^{M\times N}$ , $\hat{\mathbf{P}}=[\hat{p}_{ln}]\in\mathbb{C}^{N\times N}$ and $\hat{\mathbf{Y}}=[\hat{y}_{ln}]\in\mathbb{C}^{N\times N}$ , respectively, as

$$r_{ml}=\begin{cases}M^{-1/2} & l=(N-M)/2+m\\ 0 & else\end{cases} \tag{34}$$

$$\hat{p}_{ln}=N^{-1/2}\hat{c}_p\left[((l-n))_N\right] \tag{35}$$

$$\hat{y}_{ln}=N^{-1/2}\hat{\psi}_n\left(e^{j(-\pi+2\pi(l-1)/N)}\right) \tag{36}$$

where the notation $((k))_N$ denotes $(k \text{ modulo } N)$ . Then the matrix $\hat{\mathbf{M}}$ can be factorized as

$$\hat{\mathbf{M}}=\mathbf{R}\hat{\mathbf{P}}\hat{\mathbf{Y}} \tag{37}$$

The factorization of $\hat{\mathbf{M}}$ clearly shows the three basic components of the QuadCS system: $\hat{\mathbf{Y}}$ describes the dictionary in frequency-domain, $\hat{\mathbf{P}}$ is a circulant matrix generated by the spectrum of the spectral-spreading signal, and $\mathbf{R}$ corresponds to the operations of the bandpass filter and the low-rate sampling. Among these components, $\hat{\mathbf{Y}}$ and $\mathbf{R}$ are determinestic matrices, and $\hat{\mathbf{P}}$ is a random matrix determined by the chipping sequence $\{\varepsilon_0,\varepsilon_1,\cdots,\varepsilon_{N-1}\}$.

The matrix $\hat{\mathbf{M}}$ describes the overall actions of the QuadCS system, and establishes the relations between the baseband complex envelope and each of the QuadCS system operations. The matrix $\hat{\mathbf{M}}$ has a special place in the following analysis, and we refer it as the frequency-domain QuadCS (FD-QuadCS) matrix. In turn, the matrix $\tilde{\mathbf{M}}$ given by (17) is called as the time-domain QuadCS (TD-QuadCS) matrix.

## V. Restricted Isometry Property of the QuadCS System

This section is to show that the measurement system derived from the proposed QuadCS satisfies RIP with an appropriate setting of the sampling frequency $f_{IF}^{cs}$ or the filter bandwidth $B_{cs}$. The established RIP ensures that the QuadCS is a valid candidate for CS. The frequency-domain representation, the FD-QuadCS matrix $\hat{\mathbf{M}}$ , is exploited, because it explicitly shows the compositions of the QuadCS system.

We will prove RIP from the concentration of measure (CoM) inequalities [26~28], as done in [29~32]. The CoM inequalities have been derived for the random matrix with *i.i.d* entries [29] and other structured matrices [30~32]. However, these results can not be directly applied to the FD-QuadCS matrix $\tilde{\mathbf{M}}$ because of its special factorization.

To simplify the analysis, the dictionary matrix $\hat{\mathbf{Y}}$ in $\hat{\mathbf{M}}=\mathbf{R}\hat{\mathbf{P}}\hat{\mathbf{Y}}$ is assumed to be orthogonal such that $\hat{\mathbf{Y}}^{\mathrm{H}}\hat{\mathbf{Y}}=\mathbf{I}_N$. The rationality of the assumption in radar applications is analyzed at the end of the section. Define a circulant matrix operator $\tilde{\mathbf{C}}(\bullet):\mathbb{C}^N\to\mathbb{C}^{N\times N}$ as

$$\tilde{\mathbf{C}}(\tilde{\mathbf{u}})=\begin{bmatrix}\tilde{u}_1 & \tilde{u}_2 & \cdots & \tilde{u}_N\\ \tilde{u}_2 & & \tilde{u}_N & \tilde{u}_1\\ \vdots & & \ddots & \vdots\\ \tilde{u}_N & \tilde{u}_1 & \cdots & \tilde{u}_{N-1}\end{bmatrix} \tag{38}$$

where $\tilde{\mathbf{u}}=[\tilde{u}_1,\tilde{u}_2,\cdots,\tilde{u}_N]^{\mathrm{T}}\in\mathbb{C}^N$. The CoM inequality for the FD-QuadCS matrix $\hat{\mathbf{M}}$ is formulated in the Lemma 1.

**Lemma 1**: *Let $\tilde{\mathbf{v}}\in\mathbb{C}^N$ be fixed. Then for $\sigma_{\tilde{\mathbf{v}}}=\left\|\tilde{\mathbf{C}}\left(\hat{\mathbf{Y}}\tilde{\mathbf{v}}\right)\right\|_2$ and $M\geq 64\sigma_{\tilde{\mathbf{v}}}^2$,*

$$\mathbb{P}\left\{(1-\varepsilon)\|\tilde{\mathbf{v}}\|_2^2\leq\left\|\hat{\mathbf{M}}\tilde{\mathbf{v}}\right\|_2^2\leq(1+\varepsilon)\|\tilde{\mathbf{v}}\|_2^2\right\}\geq 1-4e^4\exp\left(-M\varepsilon/16\sigma_{\tilde{\mathbf{v}}}^2\right) \quad (39)$$

*where $\varepsilon\geq 64\sigma_{\tilde{\mathbf{v}}}^2/M$ is a constant.*

***Proof****: See Appendix A.*

It is seen that the energy of the compressive measurement $\hat{\mathbf{s}}_{cs}=\hat{\mathbf{M}}\tilde{\mathbf{v}}$ concentrate around the energy of $\tilde{\mathbf{v}}$ with a tail probability bound that decays exponentially in the dimension $M$. The probability bound is a function of the signal $\tilde{\mathbf{v}}$ and thus, the bound is non-uniform and varies with $\tilde{\mathbf{v}}$. Different from the classic CoM inequalities [26], (39) only establishes the probability bound for $\varepsilon\in\left[64\sigma_{\tilde{\mathbf{v}}}^2/M,1\right]$, which implicitly requires that $M>64\sigma_{\tilde{\mathbf{v}}}^2$.

With the CoM, we can conduct the RIP analysis. We will first apply (39) to derive the probability bound for any signal in a fixed $K$-dimensional subspace and then extend the result to all possible $K$-dimensional signals. Specifically, given $\Lambda\subset\{1,2,\cdots,N\}$ as an index set with $|\Lambda|\leq K$, let us denote $\tilde{\mathbf{V}}_\Lambda$ as the set of all vectors in $\mathbb{C}^N$ that are zero out of $\Lambda$. For a fixed $K$-dimensional subspace $\tilde{\mathbf{V}}_\Lambda$, Lemma 2 establishes the probability bound that $(1-\varepsilon)\|\tilde{\mathbf{v}}\|_2^2\leq\left\|\hat{\mathbf{M}}\tilde{\mathbf{v}}\right\|_2^2\leq(1+\varepsilon)\|\tilde{\mathbf{v}}\|_2^2$ holds for any $\tilde{\mathbf{v}}\in\tilde{\mathbf{V}}_\Lambda$.

**Lemma 2**: *Given an index set $\Lambda$ with $|\Lambda|\leq K$ and for any vector $\tilde{\mathbf{v}}\in\tilde{\mathbf{V}}_\Lambda$,*

$$\mathbb{P}\left\{(1-\varepsilon)\|\tilde{\mathbf{v}}\|_2^2\leq\left\|\hat{\mathbf{M}}\tilde{\mathbf{v}}\right\|_2^2\leq(1+\varepsilon)\|\tilde{\mathbf{v}}\|_2^2\right\}\geq 1-8e^4K^2\exp\left(-M\varepsilon/32K\right) \quad (40)$$

*where $\varepsilon>128K/M$ is a constant.*

***Proof****: See Appendix B.*

Upon Lemma 2, we can quantify the FD-QuadCS matrix $\hat{\mathbf{M}}$ to derive the RIP condition.

**Theorem** *The matrix $\hat{\mathbf{M}}$ has restricted isometry constant $\delta_K$ with probability exceeding $1-\eta$, if it satisfies*

$$M\geq 32\delta_K^{-1}K\left(K\log(eN/K)+2\log K+\log\left(\eta^{-1}\right)+6.08\right) \quad (41)$$

**Proof**: It should be noted that there will be $C_N^K$ such $K$-dimensional subspaces in $\mathbb{C}^N$. Then by Lemma 2, we know that $(1-\delta_K)\|\tilde{\mathbf{v}}\|_2^2\leq\left\|\hat{\mathbf{M}}\tilde{\mathbf{v}}\right\|_2^2\leq(1+\delta_K)\|\tilde{\mathbf{v}}\|_2^2$ holds for any $K$-sparse vector $\tilde{\mathbf{v}}\in\mathbb{C}^N$ with the probability no less than $1-8e^4K^2C_N^K\exp\left(-M\delta_K/32K\right)$.

Since

$$C_N^K=\frac{N(N-1)\cdots(N-K+1)}{K!}\leq\frac{N^K}{K!}\leq\left(\frac{eN}{K}\right)^K \quad (42)$$

Then we have

$$8e^4K^2C_N^K\exp\left(-M\delta_K/32K\right)\leq 8e^4K^{2-K}(eN)^K\exp\left(-M\delta_K/32K\right) \quad (43)$$

Thus $\eta\geq 8e^4K^{2-K}(eN)^K\exp\left(-M\delta_K/32K\right)$. We can derive (41) by some simplifications.■

The Theorem demonstrates that the number of compressive measurements is closely related to the sparsity $K$ and the signal dimension $N$ to guarantee the RIP of $\hat{\mathbf{M}}$. In general, $M=\mathcal{O}\left(K^2\log(N/K)\right)$ measurements are required for the QuadCS system to satisfy the RIP. It is well-known that for the $K$-sparse coefficient vector $\tilde{\mathbf{v}}$, $\mathcal{O}\left(K\log(N/K)\right)$ measurements are sufficient to establish the RIP of order $K$ [34]. Thus, the measurements $M$ provided by the Theorem are sub-optimal. Our result coincides with the RIP condition for the random Toeplitz or circulant matrix [30, 31]. The reason may be that the FD-QuadCS matrix $\hat{\mathbf{M}}$ also includes a random circulant matrix $\hat{\mathbf{P}}$ as its components. However, as we will demonstrate in Section VI, the theoretical result given by the Theorem is overestimated. In fact, the simulation results demonstrate that $M=\mathcal{O}\left(K\log(N/K)\right)$ is enough to guarantee the probability of successful reconstruction as high as 99%.

With the Theorem, we can establish the relation among the compressive bandwidth, the signal bandwidth and the observation interval.

***Corollary 1****: The QuadCS system satisfies the RIP with probability exceeding $1-\eta$, if the bandwidth $B_{cs}$ satisfies*

$$B_{cs}\geq 32\delta_K^{-1}(K/T)\left(K\log(eBT/K)+2\log K+\log\left(\eta^{-1}\right)+6.08\right) \quad (44)$$

*where $\delta_K$ is the restricted isometry constant.*

***Proof****: Inserting the relations $N=BT$ and $M=B_{cs}T$ into* (41)*, we have* (44).■

It is seen that the QuadCS system with the compressive bandwidth $B_{cs}=\mathcal{O}\left(\left(K^2/T\right)\log(BT/K)\right)$ is sufficient for high-probability recovery of the sparse baseband signals with the bandwidth $B$. Different from the traditional quadrature sampling, the sampling rate by the QuadCS system depends on the observation interval and the target number. In general, the rate is proportional to the target number and inversely proportional to the observation interval.

We now examine the rationality of the assumption on the orthogonal dictionary matrix $\hat{\mathbf{Y}}$. As given in (36), the elements of $\hat{\mathbf{Y}}$ are generated by sampling DTFT of the Nyquist sampling sequence $\tilde{\psi}_n[l]=\tilde{\psi}_n\left(lT_{nyq}\right)$. Noting that $\tilde{\psi}_n(t)=\tilde{s}_0(t-n\tau_0)$, we have

$$\begin{aligned}\hat{y}_{ln}&=N^{-1/2}\hat{\psi}_n\left(e^{j\omega}\right)\Big|_{\omega=-\pi+2\pi(l-1)/N}\\&=N^{-1/2}\hat{s}_0\left(e^{j\omega}\right)e^{-jn\omega}\Big|_{\omega=-\pi+2\pi(l-1)/N}\\&=N^{-1/2}\hat{s}_0\left(e^{j(-\pi+2\pi(l-1)/N)}\right)e^{-jn(-\pi+2\pi(l-1)/N)}\end{aligned} \quad (45)$$

where $\hat{s}_0\left(e^{j\omega}\right)$ denotes the DTFT of the Nyquist sampling sequence $\tilde{s}_0[l]=\tilde{s}_0\left(lT_{nyq}\right)$. Define $\hat{\mathbf{S}}_0=diag\left\{\hat{s}_0\left(e^{-j\pi}\right),\hat{s}_0\left(e^{j(-\pi+2\pi/N)}\right),\cdots,\hat{s}_0\left(e^{j(-\pi+2\pi(N-1)/N)}\right)\right\}$ and $\hat{\mathbf{T}}=\left[\hat{T}_{ln}\right]\in\mathbb{C}^{N\times N}$ with $\hat{T}_{ln}=N^{-1/2}e^{-j(-\pi+2\pi(l-1)/N)n}$. We can further factorize $\hat{\mathbf{Y}}$ as

$$\hat{\mathbf{Y}}=\hat{\mathbf{S}}_0\hat{\mathbf{T}} \quad (46)$$

Since $\hat{\mathbf{T}}^{\mathrm{H}}\hat{\mathbf{T}}=\mathbf{I}_N$, it is required that $\left(\hat{\mathbf{S}}_0\right)^{\mathrm{H}}\hat{\mathbf{S}}_0=\mathbf{I}_N$ to satisfy $\hat{\mathbf{Y}}^{\mathrm{H}}\hat{\mathbf{Y}}=\mathbf{I}_N$. This is equivalent to $\left|\hat{s}_0\left(e^{-j(-\pi+2\pi l/N)\pi}\right)\right|=1$ for $l=0,1,\cdots,N-1$, i.e., the complex baseband signal $\tilde{s}_0(t)$ having flat spectrum. In radar applications, the radar waveforms are usually designed to have the flat and wide spectrum to achieve the high range resolution. For example, the spectrum of the linear frequency modulated (LFM) signal [35] is almost flat. For phase-coded signal [35], the code can be optimized to make $\hat{\mathbf{Y}}$ orthogonal or approximately orthogonal. Thus, the assumption of the orthogonality of $\hat{\mathbf{Y}}$ is reasonable. The performance of the QuadCS system for the LFM signal and the phase-coded signal will be demonstrated in Section VI.

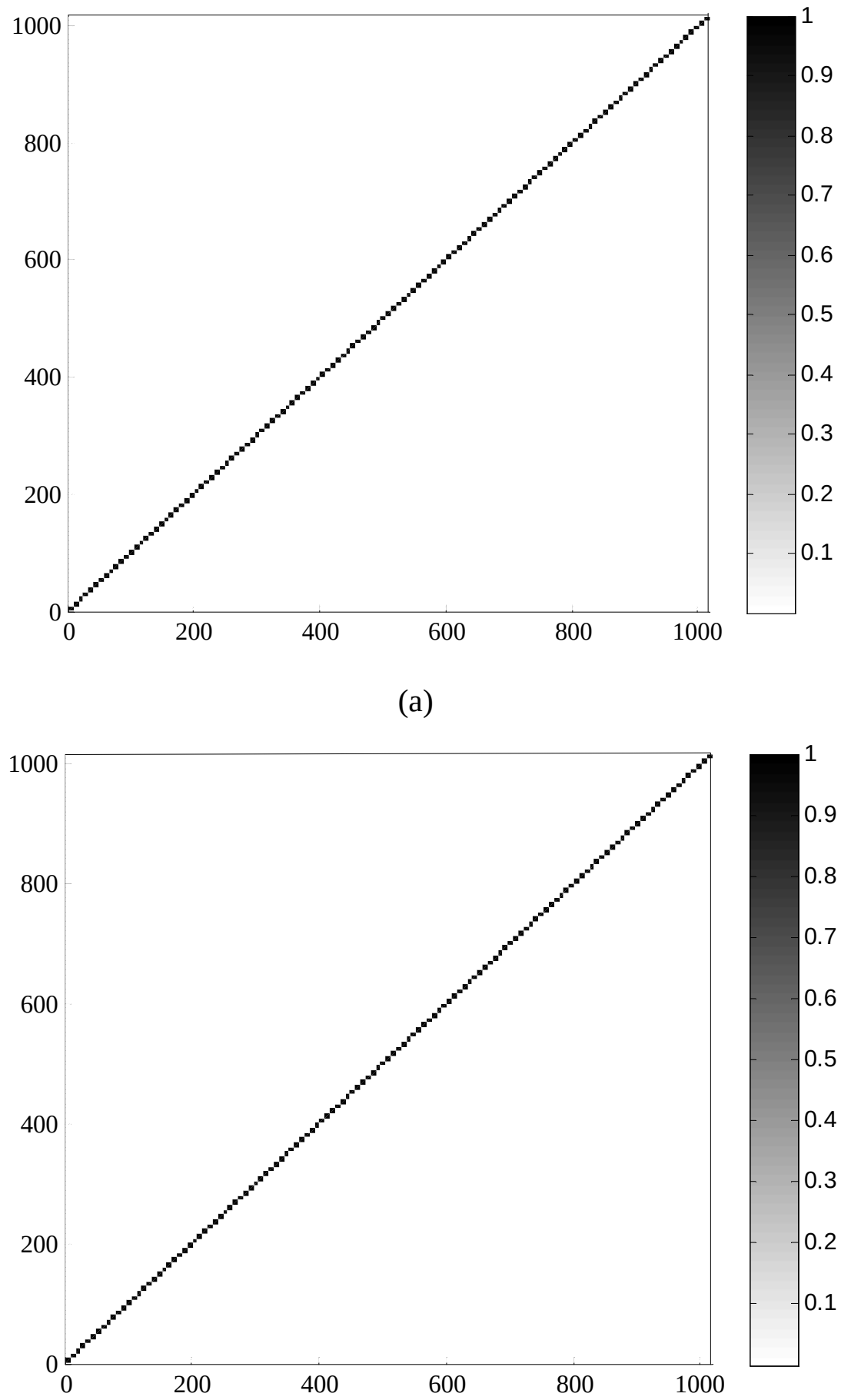

(a)

(b)

Fig.2 The Gram matrix of the waveform-matched dictionary. (a) LFM signals, (b) Phase-coded signals.

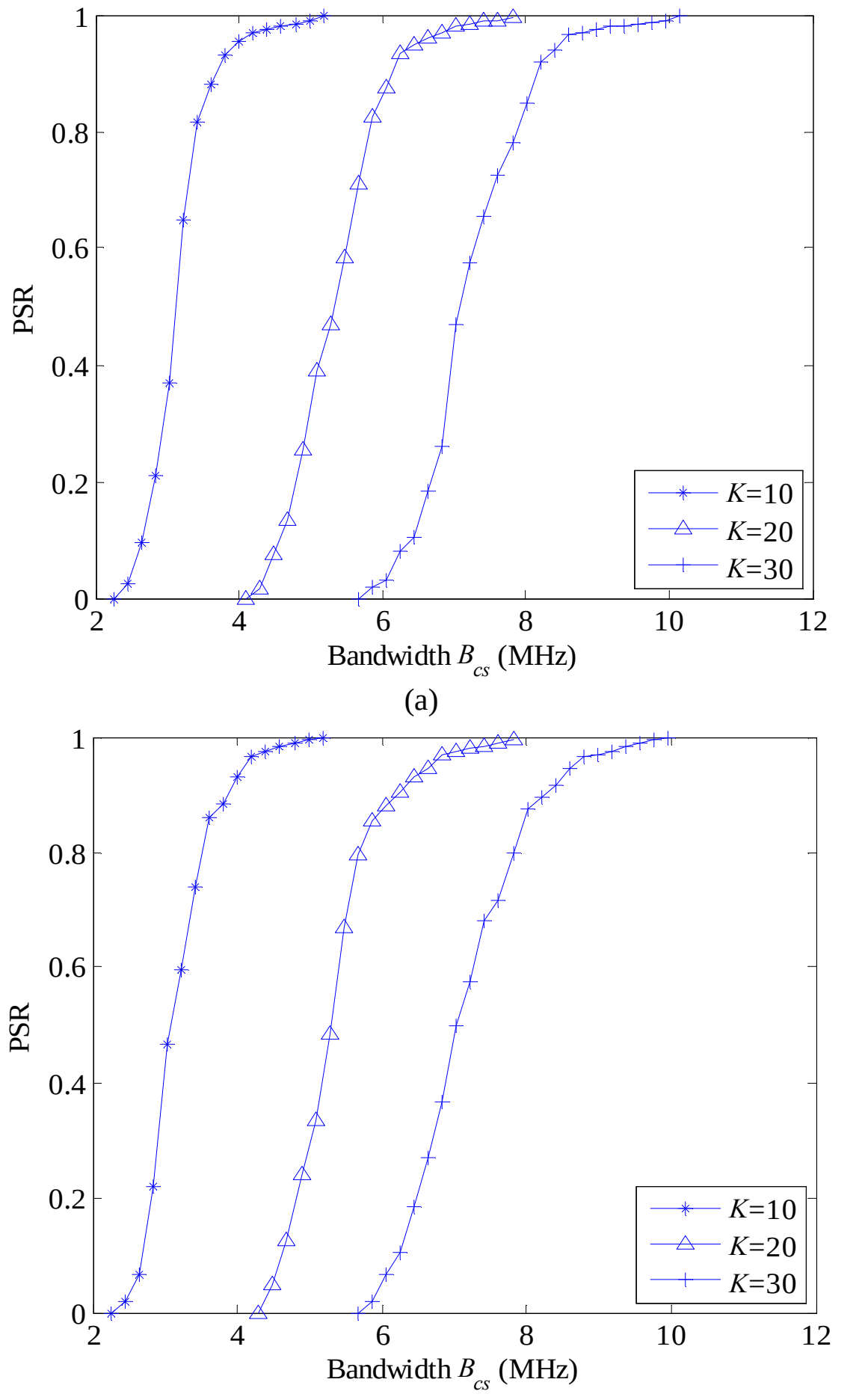

(a)

(b)

Fig.3 PSR versus bandwidth $B_{cs}$ for different sparsity levels $K$ ( $B = 100MHz$ , $T = 20.48\mu s$ ).(a) LFM signals, (b) Phase-coded signals.

## VI. Simulations

In this section, we examine various aspects of the QuadCS system through several simulation experiments. First, for the noise-free signal, we determine the minimum bandwidth $B_{cs}$ that is necessary to identify the $K$ -sparse signal with the bandwidth $B$ . Then, for the noisy signal, we demonstrate the SNR performance and the amplitude and/or phase reconstruction errors of the QuadCS system. Finally, we evaluate the performance of the QuadCS system in a more practical scenario where the time-delays of the radar echoes are randomly distributed in the observation interval, rather than located on a subset of the Nyquist sampling grids.

### A. Simulation Settings

We consider two types of waveforms which are widely used in radars:

- Linear frequency modulated (LFM) signal

The complex baseband signal $\tilde{s}_0(t)$ of a LFM waveform is represented as

$$\tilde{s}_0(t) = \text{rect}\left(\left(t - T_p/2\right)/T_p\right)\exp\left(j\pi\mu\left(t - T_p/2\right)^2\right) \tag{47}$$

where $T_p$ is the pulse width, $\mu = B/T_p$ is the chirp rate, $B$ is the bandwidth of the LFM signal, respectively, and $\text{rect}(t/T_p)$ represents a rectangular pulse

$$\text{rect}(t/T_p) = \begin{cases} 1 & -T_p/2 \le t \le T_p/2 \\ 0 & else \end{cases} \tag{48}$$

- Phase-coded signal

The phase-coded signal with the bandwidth $B$ divides the $T_p$ -width pulse into a number of bits of identical duration $T_b = 1/B$ , and each bit is coded with a different phase value. The complex baseband signal $\tilde{s}_0(t)$ of the phase-coded signal is given by

$$\tilde{s}_0(t) = \sum_{m=1}^{M_b} \tilde{b}_m \text{rect}\left(\left(t - (m-1)T_b\right)/T_b\right) \tag{49}$$

where $M_b = T_p/T_b$ is the number of bits, $\tilde{b}_m = \exp(j\phi_m)$ and the set of $M_b$ phases $\{\phi_1, \phi_2, \cdots, \phi_{M_b}\}$ is the phase code associated with $\tilde{s}_0(t)$ . For different radar applications, the different phase codes have been designed. The Zadoff-Chu code [36, 37] is assumed in the simulation.

In the simulation experiments, the signal parameters are set as $B = 100MHz$ , $T_p = 10.24\mu s$ and $T = 20.48\mu s$ . The targets are assumed to exist in the time interval $(0, T - T_p]$ . Then the dictionary size is set as $N=\lfloor B(T-T_p)\rfloor$ [1]. For the $k$ -th target, the time-delay $t_k$ is randomly chosen from the set $\{1/B,\cdots,N/B\}$ , the gain coefficient $\nu_k$ and the phase offset $\varphi_k$ are uniformly distributed between $(0,1]$ and $(0,2\pi]$ , respectively. The IF frequency $f_0$ is set between $400MHz$ and $500MHz$ , which can be tuned to achieve the minimal sampling rate $f_{IF}^{cs}=2B_{cs}$ . Without special statements, 1000 realizations are conducted and the averaged results are given in the following.

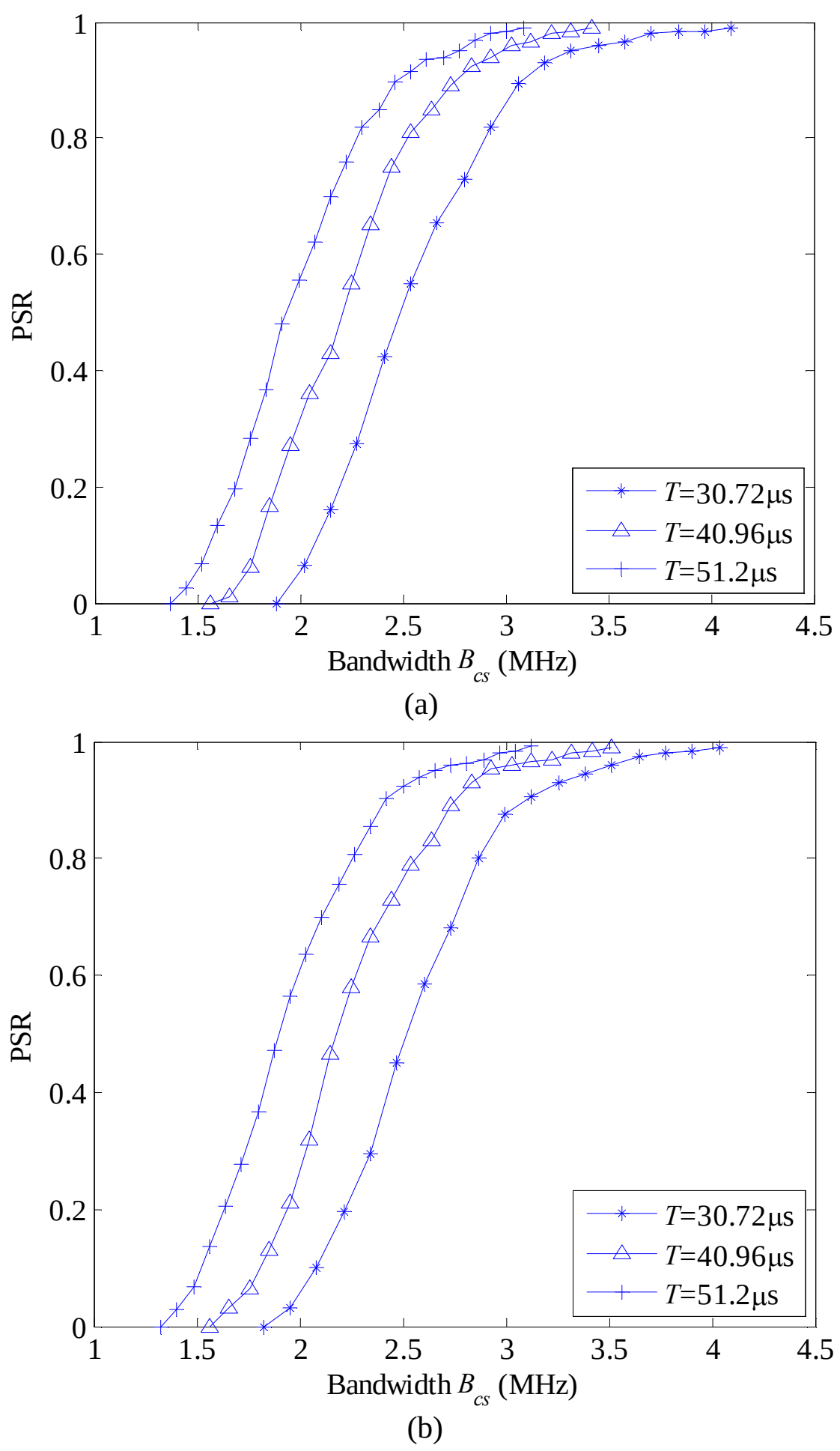

Fig.4 PSR versus bandwidth $B_{cs}$ for different observation intervals $T$ ( $B=100MHz$ , $K=10$ ). (a) LFM signals, (b) Phase-coded signals.

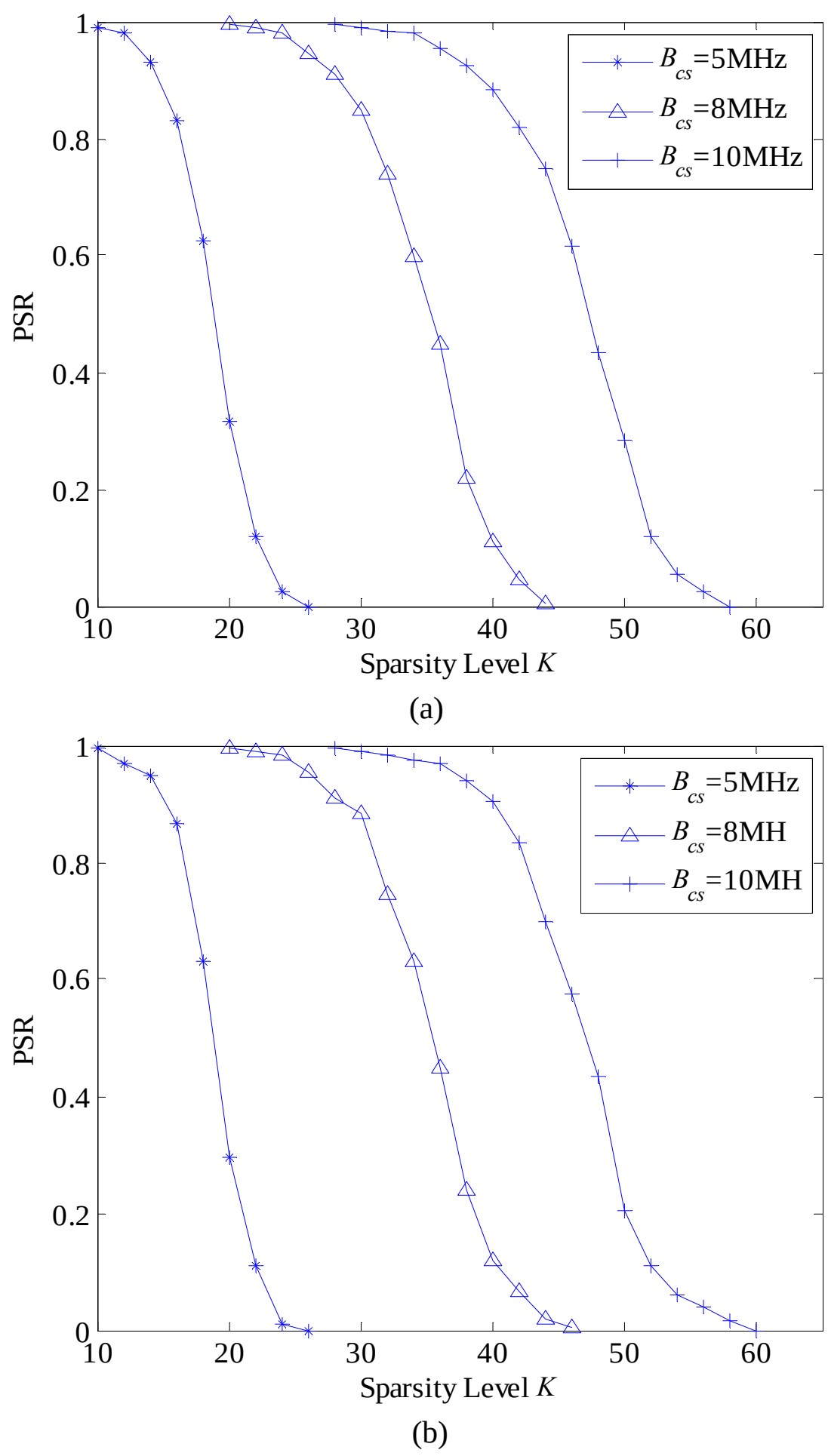

Fig.5 PSR versus sparsity level $K$ for different bandwidth $B_{cs}$ ( $B=100MHz$ , $T=20.48\mu s$ ). (a) LFM signals, (b) Phase-coded signals.

In Fig.2, we show the Gram matrix $\mathbf{G}=\hat{\mathbf{Y}}^{\mathrm{H}}\hat{\mathbf{Y}}$ of the dictionary matrix $\hat{\mathbf{Y}}$ calculated from the LFM signals and the phase-coded signals. The maximum off-diagonal elements of the Gram matrix are respectively 0.0150 and 0.0154 for the two types of signals. The Gram matrices clearly demonstrate that the dictionary matrices are approximately orthogonal.

### *B. Signal Reconstruction in the Noise-Free Case*

For the noise-free signal, we take the probability of successful reconstruction (PSR) to evaluate the QuadCS performance. A reconstruction is declared to be successful if the relative reconstruction error $E_r\le 1e-6$ with

$$E_r=\left\|\tilde{\mathbf{v}}^*-\tilde{\mathbf{v}}\right\|_2/\left\|\tilde{\mathbf{v}}\right\|_2 \tag{50}$$

where $\tilde{\mathbf{v}}^*$ is the solution of (20).

Firstly, we simulate variations of the PSR as the bandwidth $B_{cs}$ , the observation intervals $T$ , and the sparsity level $K$ . Fig. 3 shows the curves of the PSR versus the bandwidth $B_{cs}$ under different sparsity levels $K$ . Fig. 4 shows the curves of the PSR versus the bandwidth $B_{cs}$ under different observation intervals $T$ . It is seen from Fig.3 that the bandwidth $B_{cs}$ can be set to as low as 5%~10% of the bandwidth $B$ with the PSR higher than 0.99 for the assumed sparsity levels. However, the PSR degrades rapidly once the bandwidth $B_{cs}$ is lower than some threshold value. From Fig.4, we know that for a given target number, the larger the observation interval $T$ , the smaller the required bandwidth $B_{cs}$ is to achieve the given PSR. This is because more compressive measurements can be acquired for the large observation interval. Fig.5 shows the curves of the PSR versus the sparsity levels $K$ under different bandwidth $B_{cs}$ . For a fixed sparsity level $K$ , the PSR of the QuadCS system is high for large bandwidth $B_{cs}$ . For the two types of signals, the QuadCS system achieves almost the same PSR

[1] Although this assumption is slightly different from the signal model given in Section II, it will not affect the performance analysis, because $N\approx BT$ for $T\gg T_p$ .

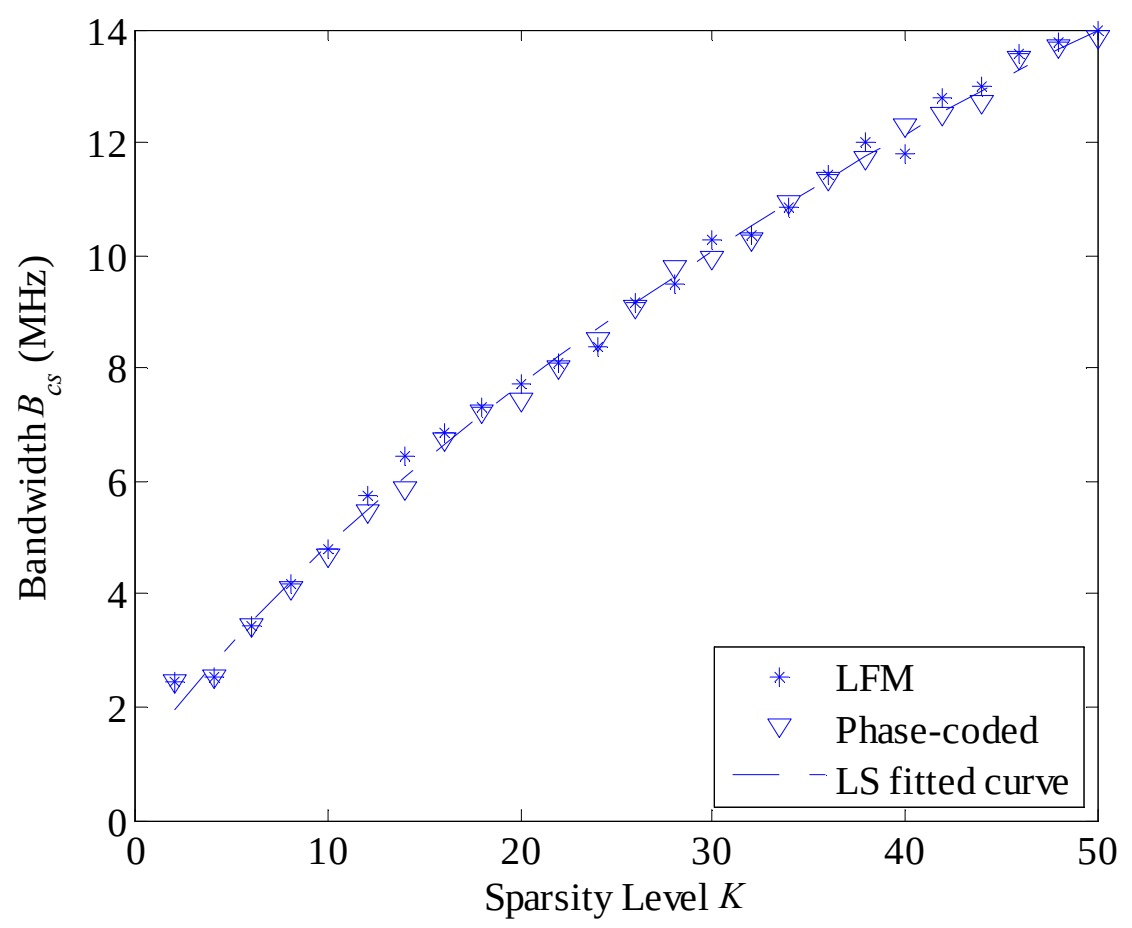

Fig.6 Bandwidth $B_{cs}$ versus sparsity level $K$ with PSR=0.99.

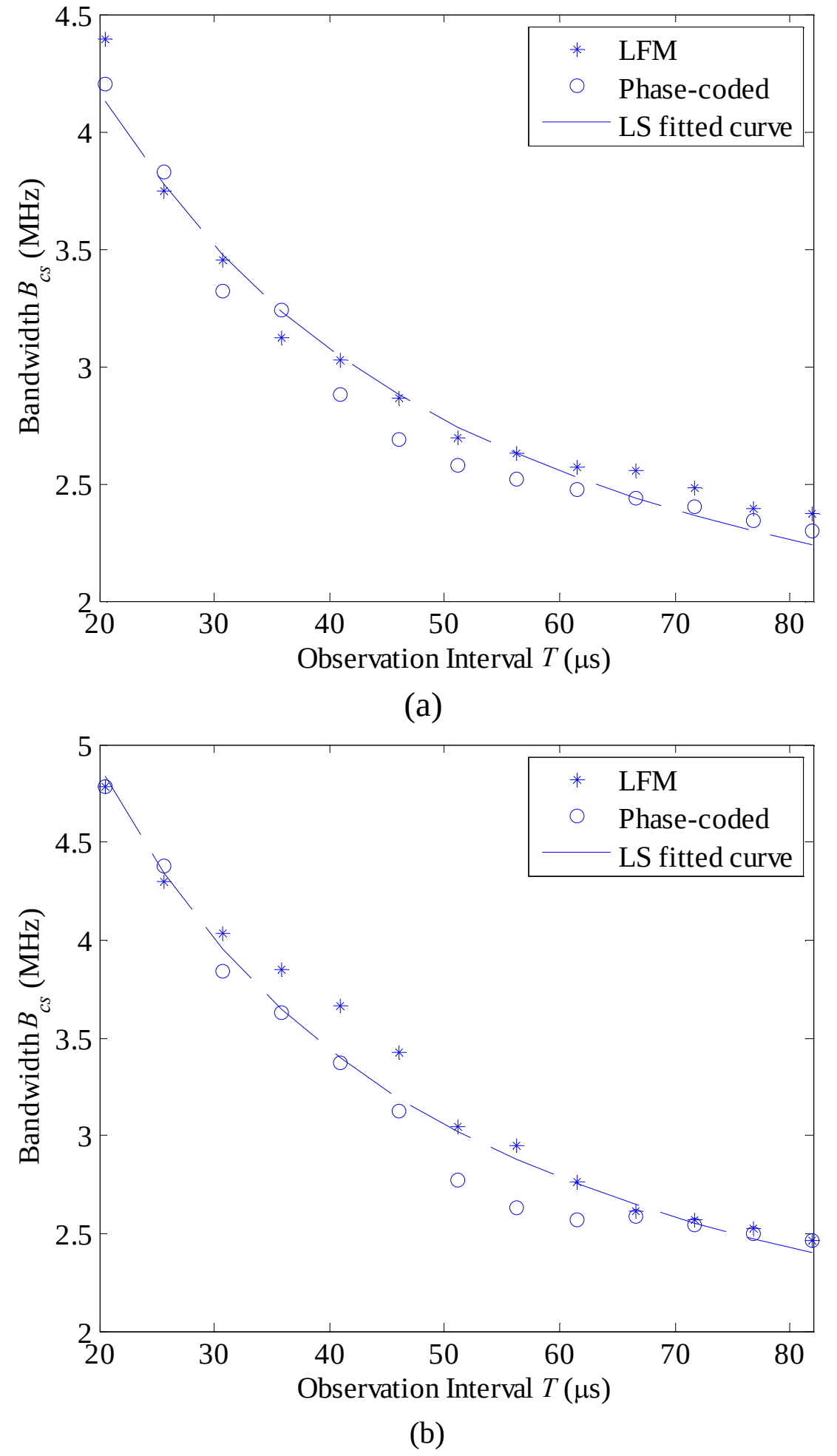

Fig.7 Bandwidth $B_{cs}$ versus observation interval $T$ with PSR=0.99. ( $B=100MHz$ , $K=10$ ). (a) $B=50MHz$ , (b) $B=100MHz$ .

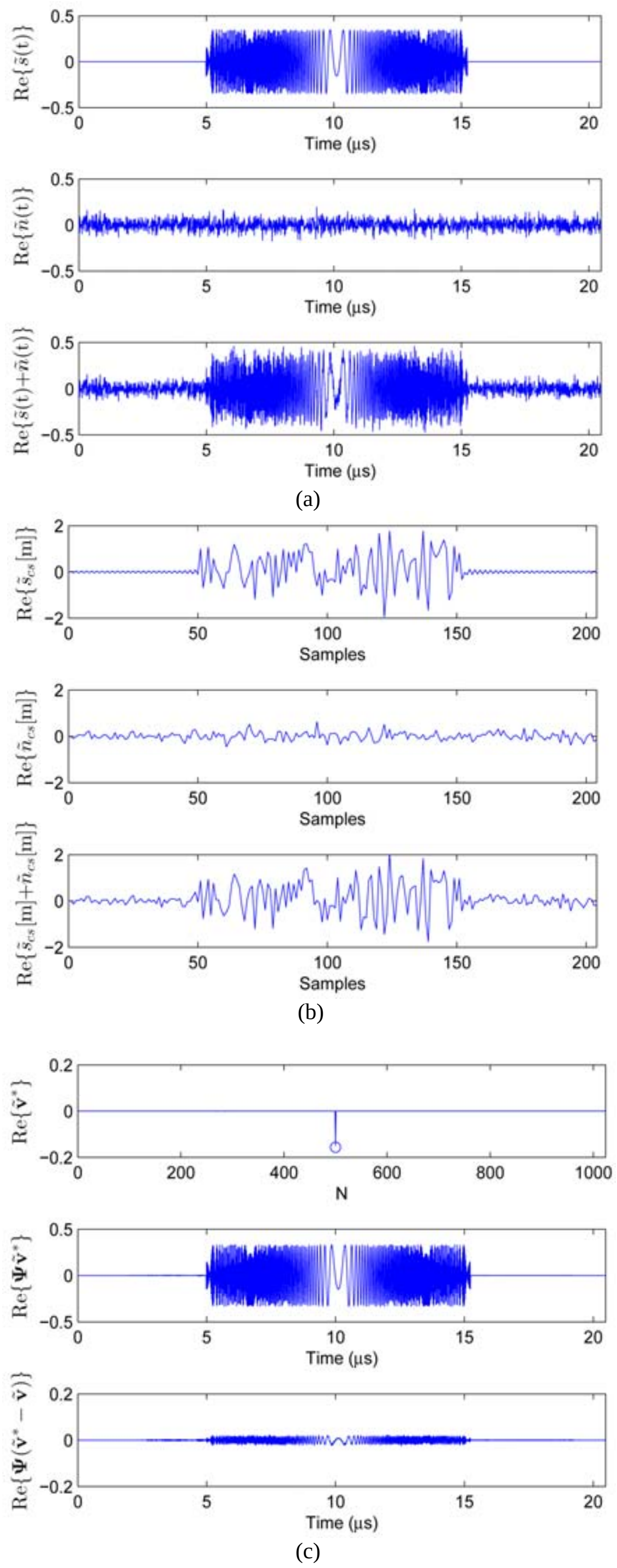

Fig.8 The input baseband components (a), compressive measurements (b) and the reconstructed signals (c) of the QuadCS system with the noisy LFM signal (ISNR=10dB, $K=1$ ).

performance.

Then we establish an empirical result of the minimum bandwidth $B_{cs}$ that is necessary to reconstruct the $K$ -sparse signals with the bandwidth $B$ and the observation time $T$ . The PSR is set as high as 99%. Fig 6 shows the experiment results on the bandwidth $B_{cs}$ versus the sparsity level $K$ , where the signal bandwidth and the observation interval are fixed as $B=100MHz$ and $T=20.48\mu s$ , respectively. The experimental results demonstrate that the required bandwidth $B_{cs}$ grows almost linearly as the sparsity level $K$ increases. The dotted line in Fig.6 shows the result of a linear regression by the least-squares (LS) fitting on the experimental data

$$B_{cs}=1.78(K/T)\log(BT/K)+0.85\times10^{6} \qquad (51)$$

Fig.7(a) and Fig.7(b) show the experiment results on the

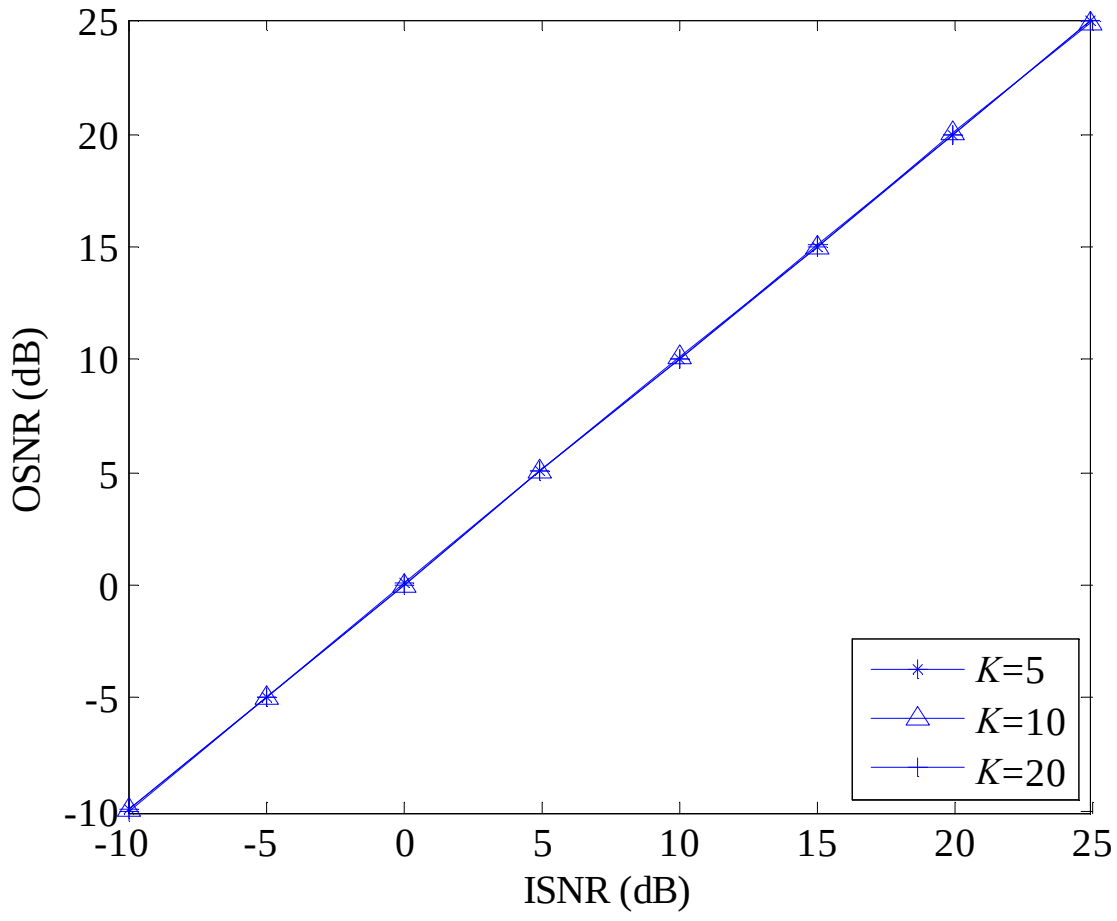

(a)

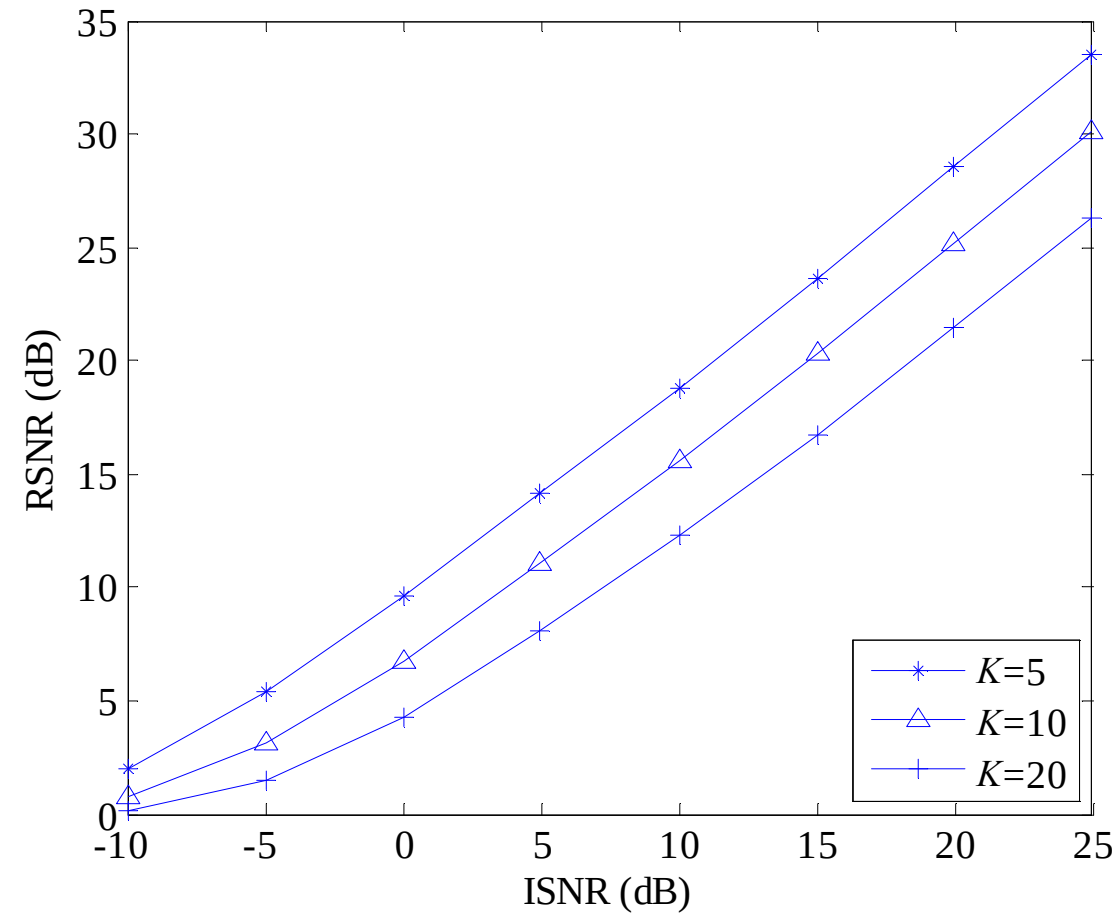

(b)

Fig. 9 The SNR performance of the QuadCS system with the noisy LFM signals ( $B_{cs}=10MHz$ ). (a) OSNR, (b) RSNR.

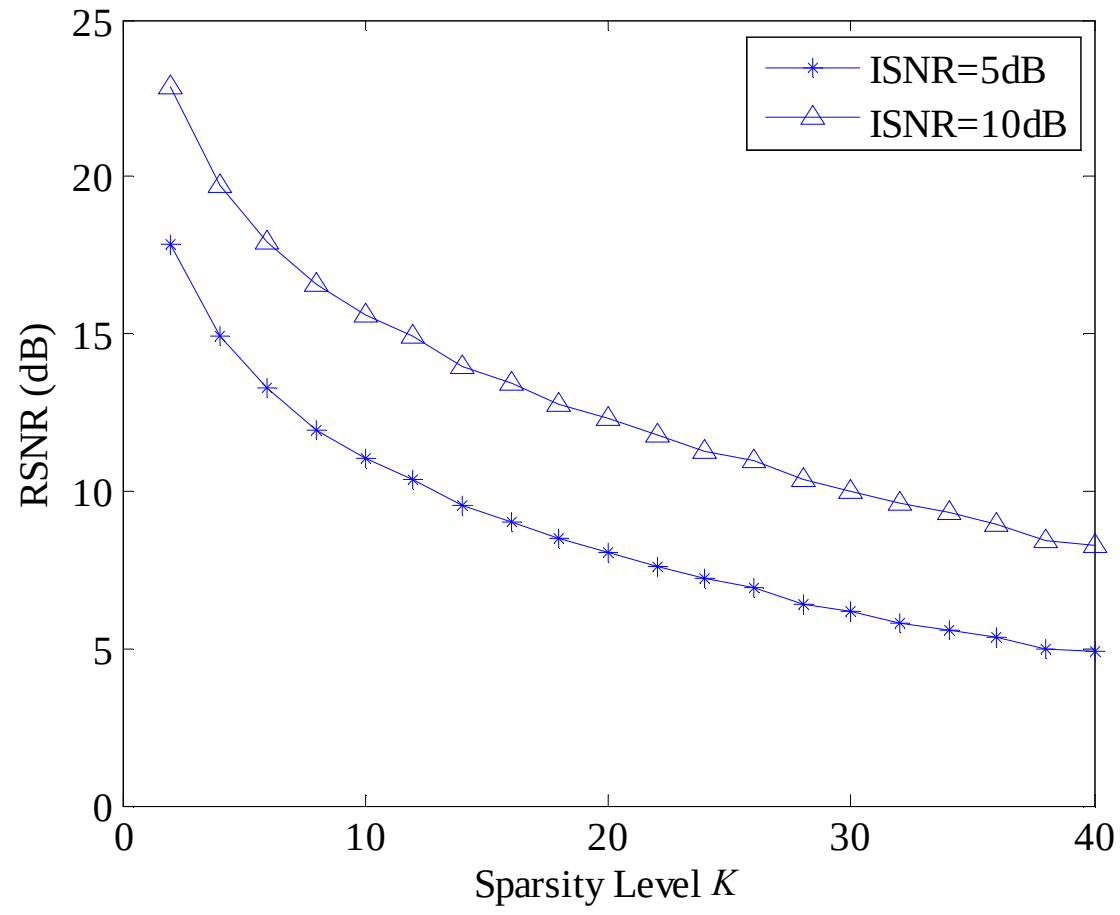

(a)

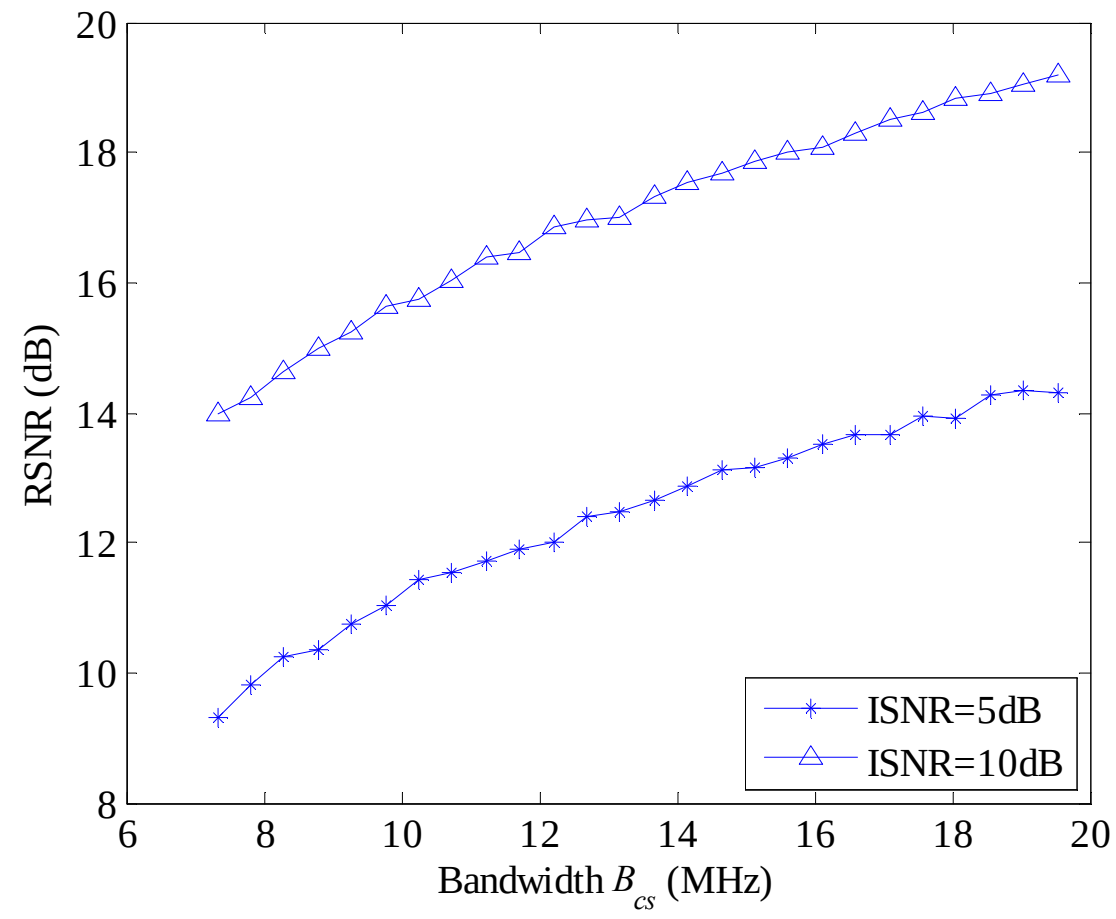

(b)

Fig.10 (a) RSNR versus the number of targets $K$ for $B_{cs}=10MHz$ , and (b) RSNR versus the bandwidth $B_{cs}$ for $K=10$ .

bandwidth $B_{cs}$ versus the observation interval $T$ with $B=50MHz$ and $B=100MHz$ , respectively. The LS fitting shows that the bandwidth $B_{cs}$ satisfies

$$B_{cs}=1.57(K/T)\log(BT/K)+1.1\times10^{6},\ B=50MHz \quad (52)$$

$$B_{cs}=1.67(K/T)\log(BT/K)+1.05\times10^{6},\ B=100MHz \quad (53)$$

These experiments suggest that the bandwidth $B_{cs}$ satisfies:

$$B_{cs}\geq1.78(K/T)\log(BT/K) \quad (54)$$

for successful reconstruction of signals with high probability. Thus, the required sampling rate of the QuadCS system grows almost linearly as the sparsity $K$ increases and logarithmically as the time-bandwidth product $BT$ increases. The fitting results also suggest that $M=\mathcal{O}(K\log(N/K))$ measurements are sufficient to guarantee the reconstruction with high PSR, where $N=BT$ is number of the Nyquist sampling. It is noted that the result is similar to those obtained for other random measurement schemes with *i.i.d* entries [29]. It is seen that the required bandwidth $B_{cs}$ of the QuadCS system can be much smaller than the theoretical results.

### C. *Signal Reconstruction in the Noise Case*

Now we simulate the reconstruction performance of the QuadCS system in the noisy case. The received IF signal is corrupted by a bandlimited, additive, white Gaussian noise $n_r(t)$ with the center frequency $f_0$, the bandwidth $B$ and the power spectral density $N_0/2$. For the noise signal, the noise measurement vector $\tilde{\mathbf{n}}_{cs}=\left[\tilde{n}_{cs}[0],\cdots,\tilde{n}_{cs}[M-1]\right]^{T}$ in (21) is given by

$$\begin{aligned}\tilde{n}_{cs}[m]&=\int_{-\infty}^{+\infty}h_{bp}(\tau)p(mT_{cs}-\tau)n_r(mT_{cs}-\tau)d\tau\\&=\int_{-\infty}^{+\infty}h_{bp}(\tau)e^{-j2\pi f_0\tau}p(mT_{cs}-\tau)\tilde{n}(mT_{cs}-\tau)d\tau\end{aligned} \quad (55)$$

where $\tilde{n}(t)$ is the baseband complex envelope of the bandpass noise $n_r(t)$. The input SNR (ISNR) is defined as

$$\text{ISNR}=\left(\int_0^T|r(t)|^2dt\Big/T\right)\Big/N_0B \quad (56)$$

To quantify the noise impact on the QuadCS system, we define two SNR's, the output SNR (OSNR) in the compressive envelope signal and the reconstructed SNR (RSNR) after the reconstruction,

$$\text{OSNR}=\|\tilde{\mathbf{s}}_{cs}\|_2^2\Big/\text{E}\left[\|\tilde{\mathbf{n}}_{cs}\|_2^2\right] \quad (57)$$

$$\text{RSNR}=\left\|\tilde{\boldsymbol{\Psi}}\tilde{\mathbf{v}}\right\|_2^2\Big/\text{E}\left[\left\|\tilde{\boldsymbol{\Psi}}\left(\tilde{\mathbf{v}}-\tilde{\mathbf{v}}^{*}\right)\right\|_2^2\right] \quad (58)$$

where $\tilde{\mathbf{v}}^{*}$ is the solution of (22). In the simulation experiments, we set the parameter $\varepsilon$ in (22) as $\varepsilon=\sqrt{NN_0B}$ .

To save the space, we show the noise performance of the QuadCS system for the LFM signals. The same conclusion can also be drawn for the phase-coded signals. Fig. 8(a) and Fig.8(b) display the waveforms of the received baseband signal and the compressive envelope. Fig.8(c) displays the reconstructed results. In the simulation, it is assumed that the received signal contains only one target signal at the time-delay $5\mu s$ with ISNR=10dB. It is seen that the target signal is well reconstructed. In fact, the reconstructed baseband signal is much "clearer" than the received one. This is because the sparsity constraint in the reconstruction removes the large amount of noise which is generally not sparse. In Fig.9, we show the OSNR and RSNR of the QuadCS system with the different ISNRs. Fig.9(a) indicates that the OSNR is almost the same as the ISNR. That is to say that the QuadCS system outputs the samples of the bandpass signals without the SNR loss even though the sampling rate is much lower than the Nyquist rate. However, the RSNR is enhanced as seen from Fig.9(b). Fig.10(a) shows the dependence of the RSNR on the sparsity levels $K$. The larger the sparsity is, the smaller the RSNR is. For example, when the sparsity level $K$ increases from 10 to 20, the RSNR decreases from 15dB to 12dB for ISNR=10dB and the RSNR decreases from 11dB to 8dB for ISNR=5dB. It seems that the RSNR decreases by 3dB for every octave increase in the sparsity level $K$. In Fig.10(b), we give the variations of the RSNR as the bandwidth $B_{cs}$. The RSNR also exhibits a 3dB loss when halving the bandwidth $B_{cs}$. The result is due to the noise-folding phenomenon [15,16], which has been proved to exist in compressive sensing.

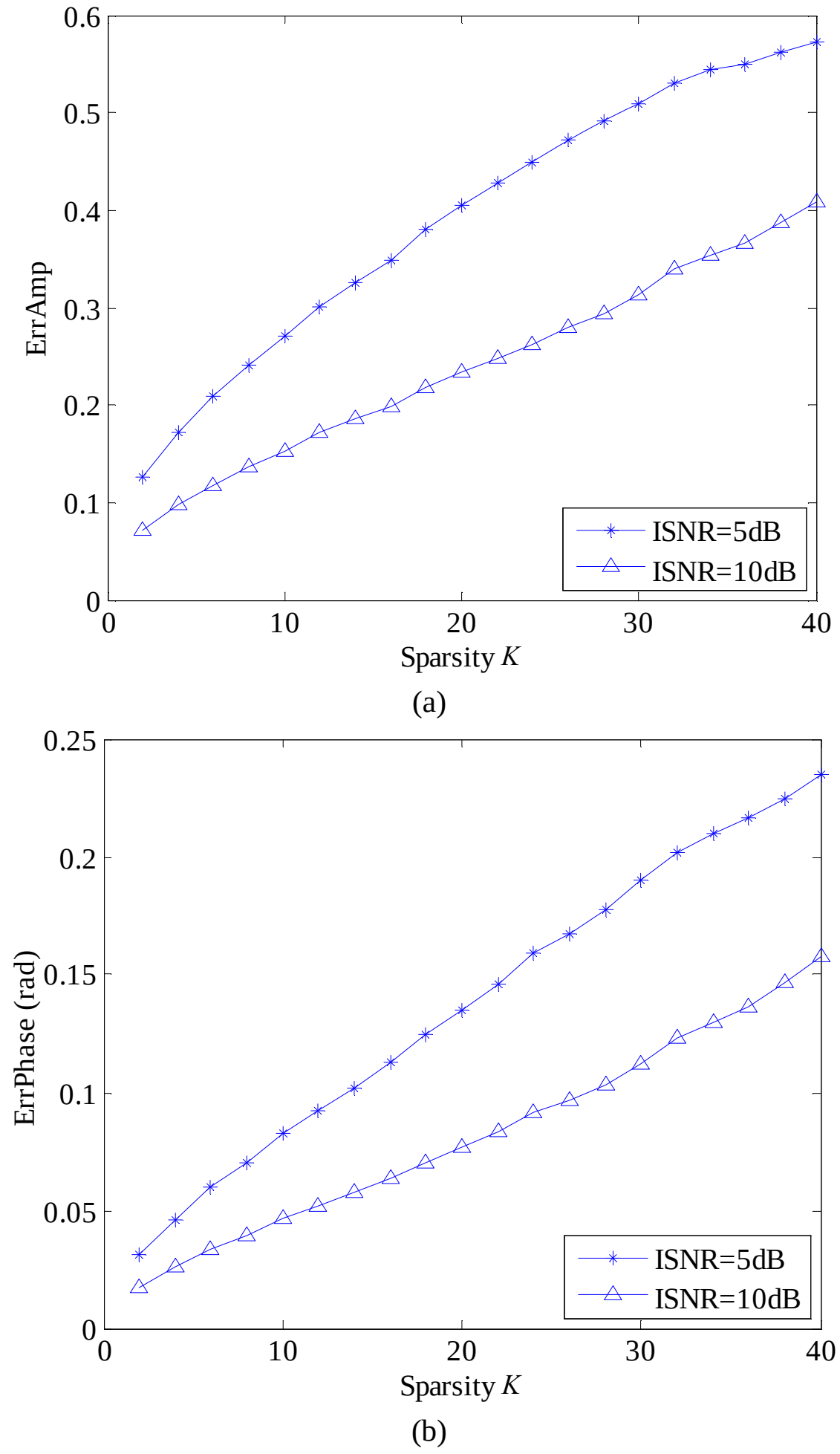

Fig.11 The amplitude and phase reconstruction errors versus the sparsity levels $K$ for $B_{cs}=10MHz$. (a) Relative amplitude error, (b) Phase error.

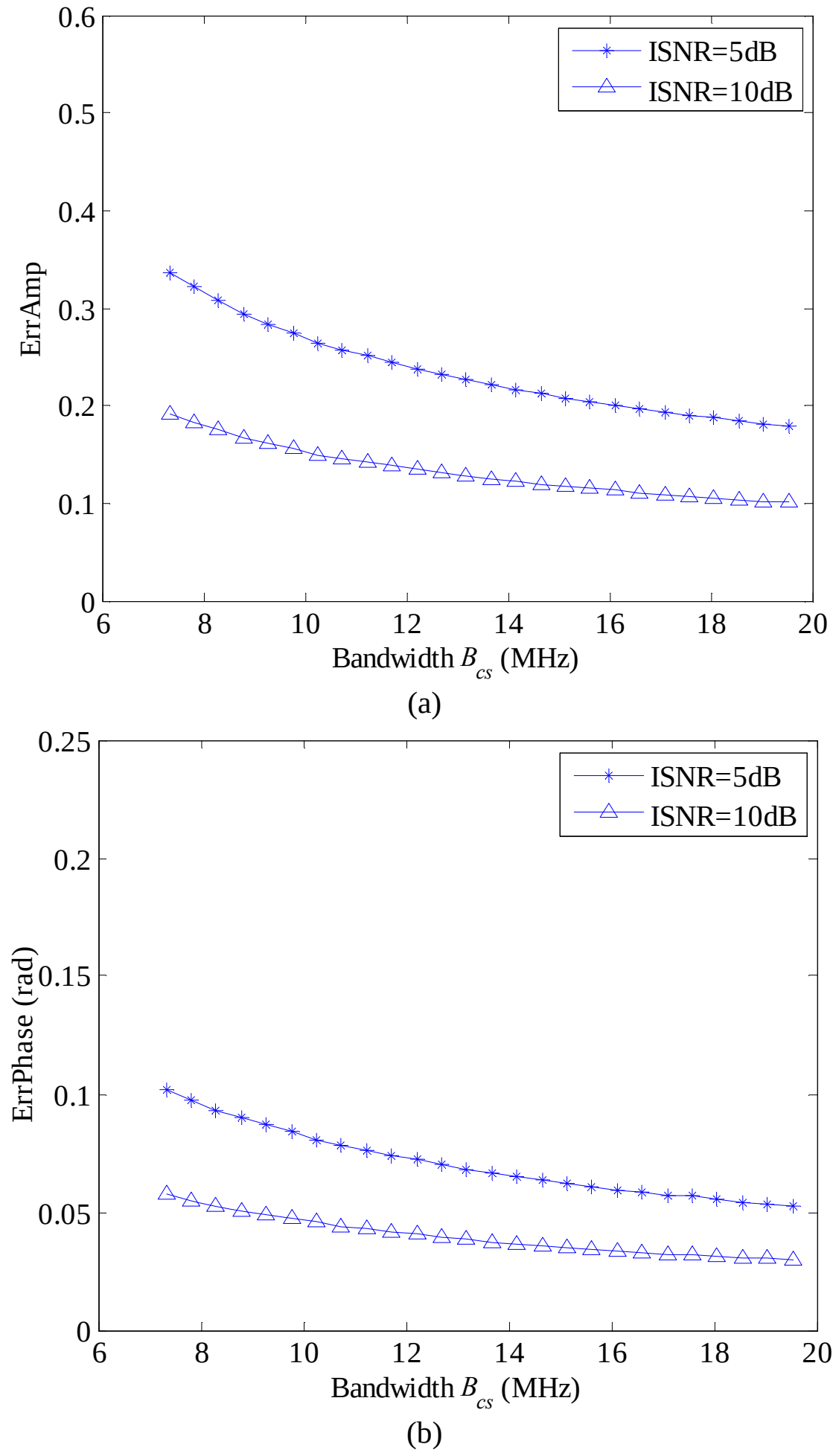

Fig.12 The amplitude and phase reconstruction errors versus the bandwidth $B_{cs}$ for $K=10$. (a) Relative amplitude error, (b) Phase error.

In radar applications, the amplitude and phase information of the radar echoes is important for the subsequent processing. We take the relative root-mean-squared amplitude error and the root-mean-squared phase error as, respectively,

$$\text{ErrAmp}=\frac{\left\| |\tilde{\boldsymbol{\Psi}}\tilde{\mathbf{v}}| - |\tilde{\boldsymbol{\Psi}}\tilde{\mathbf{v}}^*| \right\|_2}{\left\| \tilde{\boldsymbol{\Psi}}\tilde{\mathbf{v}} \right\|_2} \tag{59}$$

$$\text{ErrPhase}=\frac{1}{N}\left\| \text{Arg}\left(\tilde{\boldsymbol{\Psi}}\tilde{\mathbf{v}}\right) - \text{Arg}\left(\tilde{\boldsymbol{\Psi}}\tilde{\mathbf{v}}^*\right) \right\|_2 \tag{60}$$

to evaluate the amplitude and phase reconstruction performance. In Fig.11 and Fig.12, we give the variations of the amplitude and phase reconstruction errors (ErrAmp and ErrPhase) of the radar echoes as the sparsity level $K$ and the bandwidth $B_{cs}$, respectively. It is seen that both the ErrAmp and the ErrPhase increase as the sparsity level $K$ increases and decrease as the bandwidth $B_{cs}$ increases. As the ISNR increases, these errors decrease. The effects of the sparsity level $K$ and bandwidth $B_{cs}$ are the same as those on the RSNR in Fig.10.

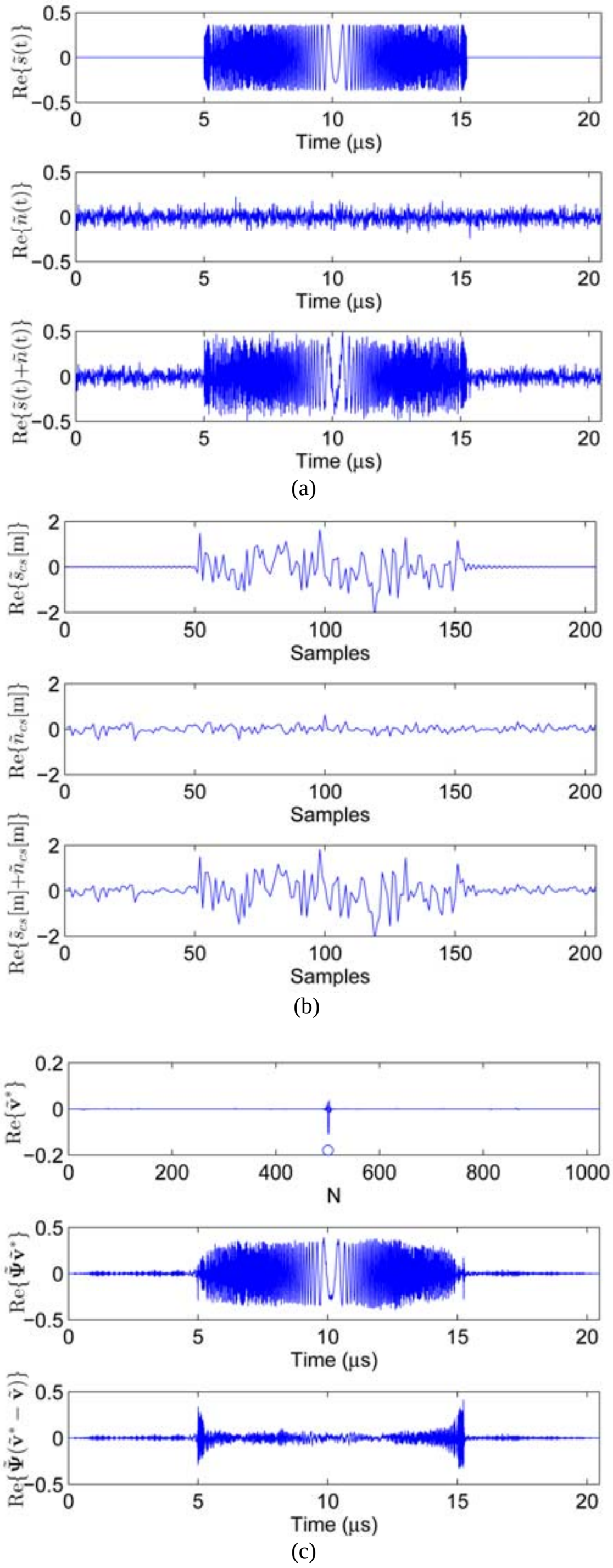

Fig.13 The input baseband components (a), compressive measurements (b) and the reconstructed signals (c) of the QuadCS system with one target echo at the time-delay $5.005\mu s$ (ISNR=10dB).

### D. Signal Reconstruction in Practical Radar Scenarios

In practical scenarios, the time-delays $\{t_k\}$ of the radar echoes may not sit exactly on the Nyquist sampling grids $\{\tau_0,\cdots,N\tau_0\}$ as that assumed in Section II, and in fact, they are randomly distributed in the observation interval $T$ . Then the real radar echoes will deviate from the assumed model. The deviation is caused by the basis mismatch [17] and will degenerate the reconstruction performance. In this subsection, we simulate the effects of the basis mismatch on the QuadCS system. To save the space, we still show the simulation results for the LFM signals.

In the simulations, the time delays $t_k$ ($1\le k\le K$ ) of the radar echoes are randomly set in the observation interval $[0.01\mu s,10.24\mu s]$ . And an additive, bandlimited, white Gaussian noise is also included in the received waveform. For the LFM signal with parameters in Part A of this section, the sampling grids consist of $\{0.01\mu s,0.02\mu s,0.03\mu s,\cdots,10.24\mu s\}$ .

In Fig.13, we display the waveforms of the received baseband signal, the compressive envelope and the reconstructed results. In the simulation, it is assumed that the received signal contains only one target signal at the time-delay $5.005\mu s$ (between the Nyquist sampling grids $5.00\mu s$ and $5.01\mu s$ ) with ISNR=10dB. In comparison with Fig.8, it is seen that the reconstructed error becomes large, especially at the rising and falling edges of the LFM pulse. It is also noted from Fig.13(c) that although the largest coefficient is found, there are some small coefficients around the largest one. That is to say, the target signal can not be well modeled by a dictionary basis because of the off-grid delay. These small coefficients correspond to the false targets in the target detection. In Fig.14, we demonstrate the RSNR of the QuadCS system as the variations of the number of targets $K$ and the bandwidth $B_{cs}$ , respectively. In comparison with Fig.10, we can find that the RSNR decreases greatly (4dB for ISNR=5dB and 6dB for ISNR=10dB) because of the basis mismatch. That is, the mismatch will introduce the large reconstruction errors. However, the RSNR still decreases by 3dB for every octave increase in the number of targets $K$ and increases by 3dB for every octave increase in the bandwidth $B_{cs}$ .

However, the RSNR loss has little effect on estimating the target positions from the point of view of the time-delay estimation. We take the *hit rate* [38] as a metric to evaluate the performance of the time-delay estimation. A "hit" is defined as a time-delay estimate which falls into the interval $[t_k-\Delta,t_k+\Delta]$, where $t_k$ is the actual time-delay and $\Delta$ is a small positive constant. Thus, the hit rate is the ratio of the number of "hit" estimates to the total number of targets. In the simulation, we take the $K$ largest nonzero coefficients falling in the corresponding interval $[t_k-\Delta,t_k+\Delta]$ as a hit. The $\Delta$ is set as 3 times the Nyquist sampling grid. The hit rate performance under different sparsity levels and bandwidth is shown in Fig.15. As a comparison, the hit rate for the radar echoes with the time-delays sitting on the grids is also given. It is seen that the hit rate performance of the off-grid targets almost approaches to that of the on-grid targets under the different bandwidth and small number of the targets. As the number of the targets increases, the false targets increase and therefore, the performance gets poor. With these time-delay estimates, the target echo reconstruction can be further improved by using the grid refinement algorithms [39,40].

## VII. Conclusions and Discussions

As a new paradigm to generate RF receiver architectures that can capture a sparse or compressible wide-band signal

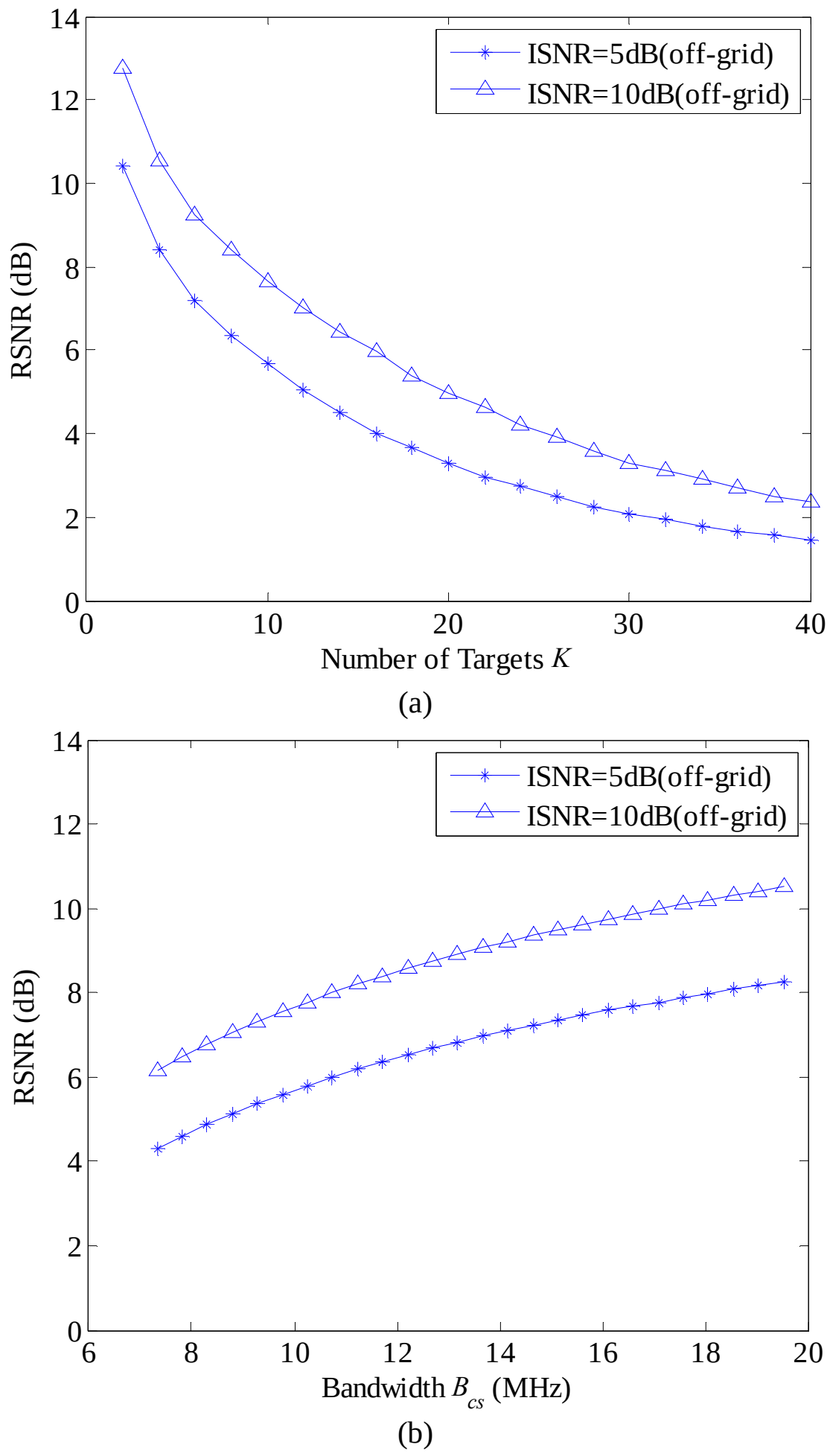

Fig.14 (a) RSNR versus the number of targets $K$ for $B_{cs}=10MHz$, and (b) RSNR versus the bandwidth $B_{cs}$ for $K=10$.

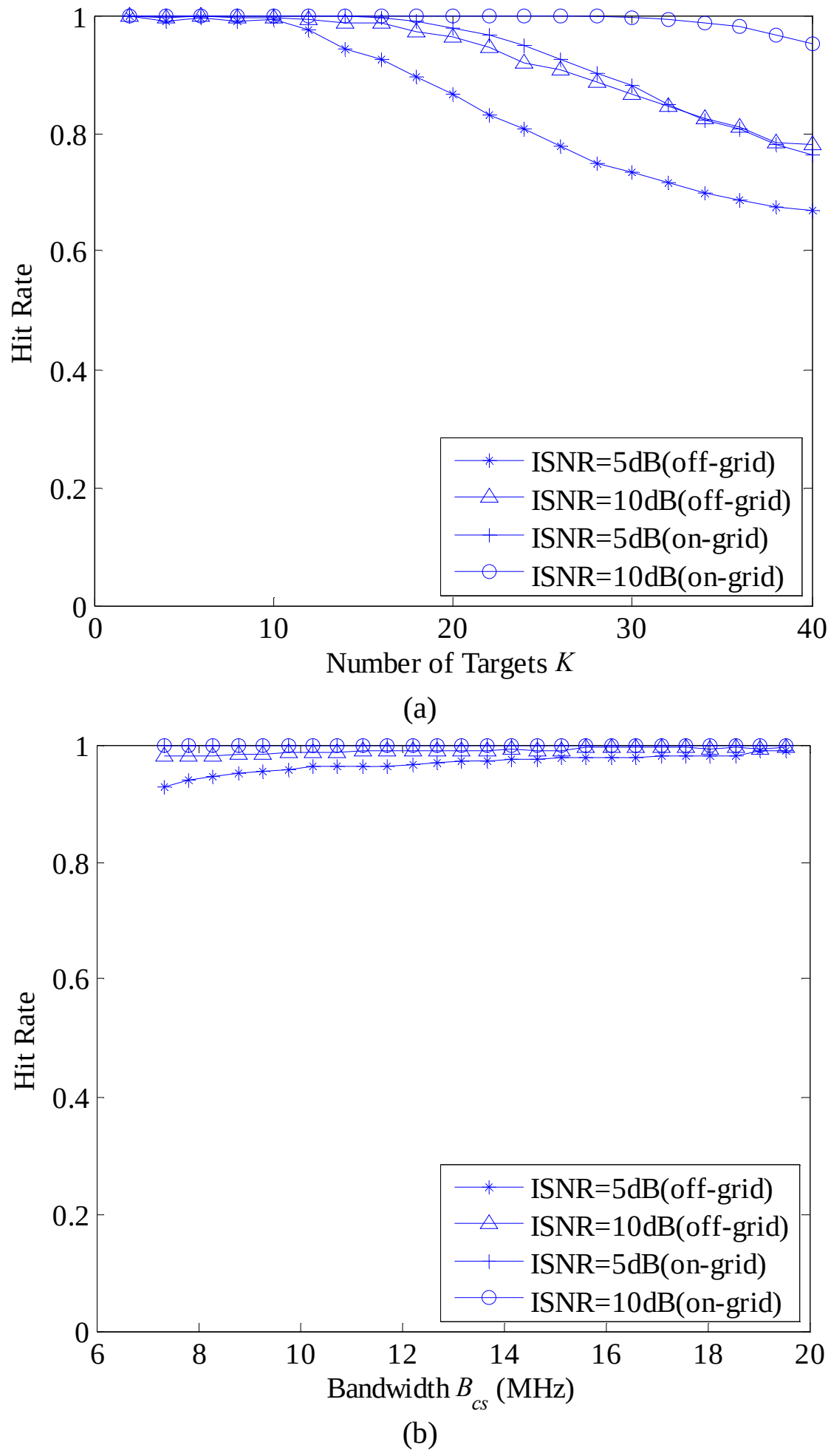

Fig.15 (a) The hit rate versus the number of targets $K$ for $B_{cs}=10MHz$, and (b) the hit rate versus the bandwidth $B_{cs}$ for $K=10$.

environment, the analog-to-information conversion has attracted wide attention in industry and academia. Different A2I techniques have been suggested with different applications and different implementation considerations. Our contribution in this area is the quadrature compressive sampling with application to the low rate sampling of radar echoes. Binding two well-developed techniques, random demodulator and quadrature sampling, the QuadCS utilizes the a priori information of the radar transmitting signals and allows a direct low-rate acquisition of the $I$ and $Q$ components, which can not be realized by any other related techniques. As we have shown, the proposed QuadCS is valid and effective for low-rate sampling of radar signals. Although the theoretical analyses and simulations are conducted for pulsed radar signals, with slight setting of the dictionary, the QuadCS is also applicable to other types of radio signals.

There are a number of interesting areas of investigation left for future research. Here we suggest three application-related aspects. One is the circuit implementation of the QuadCS system and its performance test. As we have noted, the QuadCS system is a cascade of the random demodulation and the quadrature sampling system. Then it is convenient to implement it by the off-the-shelf components. However, the calibration techniques should be developed to take into account of the clock jitter, the filter imperfections and other non-idealities. The design and circuit implementation of the QuadCS system are under investigation.

Another one is the subsequent signal processing to extract the information from the data provided by the QuadCS system. With the reconstructed signals, the information extraction can be done under the conventional processing framework. However, the clutter, interference and the basis mismatch will degenerate the reconstruction performance and thus robust reconstruction methods should be developed. The recently developed compressive signal processing [41], which directly processes the compressive measurements, provides us a new processing framework with the QuadCS system.

The third one is the reconstruction of full range data at the Nyquist rate. For large pulse repetition interval, the dictionary size $N$ will be large. Then the storage of the measurement matrix and the computational cost will be prohibitively large. This is an inherent problem in the CS non-adaptive reconstruction. The problem has gotten attention. One way to solve this problem is to employ the streaming signal reconstruction algorithms, which can recover the full range data by sequence, as done in [42,43]. Another way is to exploit the inherent structure of the dictionary to develop specific reconstruction algorithms. For the pulsed radar signals, the length of the dictionary basis is much shorter than that of the compressive measurements and then the measurement matrix will be block-banded. With the block-banded matrix, the matrix storage and the computational cost can be greatly reduced.

Acknowledgement

The authors would like to thank the anonymous reviewers for their constructive suggestions and remarks, which brought to our attention several important issues and definitely improved the paper.

Appendix

*A. Proof of Lemma 1*

The proof of Lemma 1 follows the guidelines in [26,32,33]. The basic idea is based on concentration inequalities for Lipschitz functions of random variables [26,32]. Note that $\mathrm{E}\left[\|\hat{\mathbf{M}}\tilde{\mathbf{v}}\|_2^2\right]=\|\tilde{\mathbf{v}}\|_2^2$ since $\mathrm{E}\left[\hat{\mathbf{M}}^{\mathrm{H}}\hat{\mathbf{M}}\right]=\mathbf{I}_N$. Then it is enough to show that $\|\hat{\mathbf{M}}\tilde{\mathbf{v}}\|_2^2$ concentrates around its expected value $\|\tilde{\mathbf{v}}\|_2^2$. For the linear measurement $\hat{\mathbf{M}}$, we can assume that $\|\tilde{\mathbf{v}}\|_2=1$.

Note that $\hat{\mathbf{M}}=\mathbf{R}\hat{\mathbf{P}}\hat{\mathbf{Y}}$ where $\hat{\mathbf{P}}=[\hat{p}_{ln}]\in\mathbb{C}^{N\times N}$ is a random circulant matrix with $\hat{p}_{ln}=N^{-1/2}\hat{c}_p[((l-n))_N]$, where $\hat{c}_p[k]$ is the DFT of the sequence $\{\varepsilon_0,\varepsilon_1,\cdots,\varepsilon_{N-1}\}$. Define $\mathbf{p}=[\varepsilon_0,\varepsilon_1,\cdots,\varepsilon_{N-1}]^{\mathrm{T}}$. Let us first express $\|\hat{\mathbf{M}}\tilde{\mathbf{v}}\|_2^2$ as the function of $\mathbf{p}$. With the definition of $\hat{\mathbf{P}}$ in (35), the row vector $\hat{\mathbf{p}}_m$ ($m=1,2,\cdots,N$) of $\hat{\mathbf{P}}$ can be represented as:

$$\hat{\mathbf{p}}_m=\hat{\mathbf{p}}_1\mathbf{D}^{m-1} \tag{A.1}$$

where $\mathbf{D}$ is the row shifting matrix:

$$\mathbf{D}=\begin{bmatrix} & 1 & & \\ & & \ddots & \\ & & & 1 \\ 1 & & & \end{bmatrix} \tag{A.2}$$

By (35), we have

$$\hat{\mathbf{p}}_1^{\mathrm{T}}=N^{-1/2}\mathbf{B}\tilde{\mathbf{F}}\mathbf{p} \tag{A.3}$$

where $\tilde{\mathbf{F}}=\left[\tilde{F}_{lk}\right]\in\mathbb{C}^{N\times N}$ is the $N$-dimensional discrete Fourier matrix with $\tilde{F}_{lk}=\exp\left(-j2\pi lk/N\right)$, and

$$\mathbf{B}=\begin{bmatrix} 1 & & & \\ & & & 1 \\ & & \iddots & \\ & 1 & & \end{bmatrix} \tag{A.4}$$

Define the set $\Omega=\left\{\left(N-M\right)/2+1,\left(N-M\right)/2+2,\cdots,\left(N-M\right)/2\right\}$. With the definition of $\mathbf{R}$ in (34), $\|\hat{\mathbf{M}}\tilde{\mathbf{v}}\|_2^2$ can be written as

$$\begin{aligned}\|\hat{\mathbf{M}}\tilde{\mathbf{v}}\|_2^2&=\|\mathbf{R}\hat{\mathbf{P}}\hat{\mathbf{Y}}\tilde{\mathbf{v}}\|_2^2\\&=M^{-1}\sum_{m\in\Omega}\left|\hat{\mathbf{p}}_1\mathbf{D}^{m-1}\hat{\mathbf{Y}}\tilde{\mathbf{v}}\right|^2\\&=M^{-1}\|\tilde{\mathbf{U}}_\Omega\hat{\mathbf{p}}_1^{\mathrm{T}}\|_2^2\\&=(MN)^{-1}\|\tilde{\mathbf{U}}_\Omega\mathbf{B}\tilde{\mathbf{F}}\mathbf{p}\|_2^2\end{aligned} \tag{A.5}$$

where $\tilde{\mathbf{U}}_\Omega\in\mathbb{C}^{M\times N}$ is given as

$$\tilde{\mathbf{U}}_\Omega=\begin{bmatrix}\left(\mathbf{D}^{m_1-1}\hat{\mathbf{Y}}\tilde{\mathbf{v}}\right)^{\mathrm{T}}\\\left(\mathbf{D}^{m_2-1}\hat{\mathbf{Y}}\tilde{\mathbf{v}}\right)^{\mathrm{T}}\\\vdots\\\left(\mathbf{D}^{m_M-1}\hat{\mathbf{Y}}\tilde{\mathbf{v}}\right)^{\mathrm{T}}\end{bmatrix} \tag{A.6}$$

with $m_i=\left(N-M\right)/2+i$ ($i=1,2,\cdots,M$).

Next, let $f(\mathbf{p})=\|\hat{\mathbf{M}}\tilde{\mathbf{v}}\|_2=(MN)^{-1/2}\|\tilde{\mathbf{U}}_\Omega\mathbf{B}\tilde{\mathbf{F}}\mathbf{p}\|_2$. For a fixed $\tilde{\mathbf{v}}$, we prove that $f(\mathbf{p})$ is a Lipschitz function of the random vector $\mathbf{p}$. For two independent realizations $\mathbf{p}_1$ and $\mathbf{p}_2$ of $\mathbf{p}$, we have:

$$\begin{aligned}\left|f(\mathbf{p}_1)-f(\mathbf{p}_2)\right|&=(MN)^{-1/2}\left|\|\tilde{\mathbf{U}}_\Omega\mathbf{B}\tilde{\mathbf{F}}\mathbf{p}_1\|_2-\|\tilde{\mathbf{U}}_\Omega\mathbf{B}\tilde{\mathbf{F}}\mathbf{p}_2\|_2\right|\\&\le(MN)^{-1/2}\|\tilde{\mathbf{U}}_\Omega\mathbf{B}\tilde{\mathbf{F}}(\mathbf{p}_1-\mathbf{p}_2)\|_2\\&\le(MN)^{-1/2}\|\tilde{\mathbf{U}}_\Omega\|_2\|\mathbf{B}\tilde{\mathbf{F}}\|_2\|\mathbf{p}_1-\mathbf{p}_2\|_2\\&=M^{-1/2}\|\tilde{\mathbf{U}}_\Omega\|_2\|\mathbf{p}_1-\mathbf{p}_2\|_2\end{aligned} \tag{A.7}$$

since $\|\mathbf{B}\tilde{\mathbf{F}}\|_2=N^{1/2}$. Note that $\tilde{\mathbf{U}}_\Omega$ is a submatrix of the matrix $\tilde{\mathbf{C}}\left(\hat{\mathbf{Y}}\tilde{\mathbf{v}}\right)$. Thus, $\|\tilde{\mathbf{U}}_\Omega\|_2\le\left\|\tilde{\mathbf{C}}\left(\hat{\mathbf{Y}}\tilde{\mathbf{v}}\right)\right\|_2=\sigma_{\tilde{\mathbf{v}}}$. Then $f(\mathbf{p})$ is a $M^{-1/2}\sigma_{\tilde{\mathbf{v}}}$-Lipschitz function of $\mathbf{p}$.

Finally, we use the concentration property of Lipschitz functions [26, 32] to derive the CoM inequality (39). Since each element of $\mathbf{p}$ is an *i.i.d.* Bernoulli random variable, then by the Theorem 6.6 of [33], we have the following result:

$$\mathbb{P}\left\{\left|f(\mathbf{p})-m_f\right|\ge\varepsilon\right\}\le4\exp\left(-M\varepsilon^2/16\sigma_{\tilde{\mathbf{v}}}^2\right)\qquad(\varepsilon>0) \tag{A.8}$$

where $m_f$ is the median of $f(\mathbf{p})$ satisfying

$$\mathbb{P}\left\{f(\mathbf{p})\ge m_f\right\}\le\frac{1}{2}\quad and\quad\mathbb{P}\left\{f(\mathbf{p})\le m_f\right\}\ge\frac{1}{2} \tag{A.9}$$

Since $\|\hat{\mathbf{M}}\tilde{\mathbf{v}}\|_2^2=\left(f(\mathbf{p})\right)^2$, the median of $\|\hat{\mathbf{M}}\tilde{\mathbf{v}}\|_2^2$ is equal to $m_f^2$. For a given $\varepsilon>0$, $\left|f(\mathbf{p})-m_f\right|\le\sqrt{\varepsilon}$ will result in $\left|\|\hat{\mathbf{M}}\tilde{\mathbf{v}}\|_2^2-m_f^2\right|\le\varepsilon$. Thus, we have:

$$\begin{aligned}\mathbb{P}\left\{\left|\|\hat{\mathbf{M}}\tilde{\mathbf{v}}\|_2^2-m_f^2\right|\ge\varepsilon\right\}&\le\mathbb{P}\left\{\left|f(\mathbf{p})-m_f\right|\ge\sqrt{\varepsilon}\right\}\\&\le4\exp\left(-M\varepsilon/16\sigma_{\tilde{\mathbf{v}}}^2\right)\end{aligned} \tag{A.10}$$

By the Proposition 1.8 in [26], (A.10) leads to the following result:

$$\mathbb{P}\left\{\left|\|\hat{\mathbf{M}}\tilde{\mathbf{v}}\|_2^2-1\right|\ge\varepsilon+\varepsilon_0\right\}\le4\exp\left(-M\varepsilon/16\sigma_{\tilde{\mathbf{v}}}^2\right) \tag{A.11}$$

where

$$\varepsilon_0=\int_0^{+\infty}4\exp\left(-M\mu/16\sigma_{\tilde{\mathbf{v}}}^2\right)d\mu=64\sigma_{\tilde{\mathbf{v}}}^2/M \tag{A.12}$$

Or, equivalently for $\varepsilon>64\sigma_{\tilde{\mathbf{v}}}^2/M$, we have

$$\mathbb{P}\left\{\left|\|\hat{\mathbf{M}}\tilde{\mathbf{v}}\|_2^2-1\right|\ge\varepsilon\right\}\le4e^4\exp\left(-M\varepsilon/16\sigma_{\tilde{\mathbf{v}}}^2\right) \tag{A.13}$$

As a result, for the general vector $\tilde{\mathbf{v}}\in\mathbb{C}^N$, we have

$$\mathbb{P}\left\{\left|\|\hat{\mathbf{M}}\tilde{\mathbf{v}}\|_2^2-\|\tilde{\mathbf{v}}\|_2^2\right|\ge\varepsilon\|\tilde{\mathbf{v}}\|_2^2\right\}\le4e^4\exp\left(-M\varepsilon/16\sigma_{\tilde{\mathbf{v}}}^2\right) \tag{A.14}$$

Since $0\le\varepsilon\le1$, it is required that $64\sigma_{\tilde{\mathbf{v}}}^2/M<1$. As a result, we have $M>64\sigma_{\tilde{\mathbf{v}}}^2$. ■

## B. *Proof of Lemma 2*

For the simplicity of analysis, we also focus on $\tilde{\mathbf{v}}$ with $\|\tilde{\mathbf{v}}\|_2 = 1$.

Firstly, we determine an upper bound of $\sigma_{\tilde{\mathbf{v}}}$ for any $K$-sparse $\tilde{\mathbf{v}}$. Note that $\tilde{\mathbf{C}}(\hat{\mathbf{Y}}\tilde{\mathbf{v}})$ is a circulant matrix, which can be factorized as:

$$\tilde{\mathbf{C}}(\hat{\mathbf{Y}}\tilde{\mathbf{v}}) = N^{-1}\tilde{\mathbf{F}}\,diag\left(N^{-1/2}\tilde{\mathbf{F}}^{\mathrm{H}}\hat{\mathbf{Y}}\tilde{\mathbf{v}}\right)\tilde{\mathbf{F}}^{\mathrm{H}} \quad \text{(B.1)}$$

Thus, $\sigma_{\tilde{\mathbf{v}}}$ equals to the largest amplitude of $N^{-1/2}\tilde{\mathbf{F}}^{\mathrm{H}}\hat{\mathbf{Y}}\tilde{\mathbf{v}}$, *i.e.*,

$$\begin{aligned} \sigma_{\tilde{\mathbf{v}}} &= N^{-1/2}\left\|\tilde{\mathbf{F}}^{\mathrm{H}}\hat{\mathbf{Y}}\tilde{\mathbf{v}}\right\|_\infty \\ &\le N^{-1/2}\sum_{i\in\Lambda}|\tilde{v}_i|\left\|\tilde{\mathbf{F}}^{\mathrm{H}}\hat{\mathbf{y}}_i\right\|_\infty \\ &\le K^{1/2}N^{-1/2}\sqrt{\sum_{i\in\Lambda}|\tilde{v}_i|^2}\max_i\left\|\tilde{\mathbf{F}}^{\mathrm{H}}\hat{\mathbf{y}}_i\right\|_\infty = K^{1/2}N^{-1/2}\max_i\left\|\tilde{\mathbf{F}}^{\mathrm{H}}\hat{\mathbf{y}}_i\right\|_\infty \end{aligned} \quad \text{(B.2)}$$

where $\hat{\mathbf{y}}_i$ denotes the $i$ th column of $\hat{\mathbf{Y}}$. Since $\hat{\mathbf{Y}}$ is orthogonormal, $\|\hat{\mathbf{y}}_i\|_2 = 1$ and then

$$\max_i\left\|\tilde{\mathbf{F}}^{\mathrm{H}}\hat{\mathbf{y}}_i\right\|_\infty \le N^{1/2} \quad \text{(B.3)}$$

Thus,

$$\sigma_{\tilde{\mathbf{v}}} \le K^{1/2} \quad \text{(B.4)}$$

Then by Lemma 1, for any fixed $K$-sparse $\tilde{\mathbf{v}}$, we have

$$\mathbb{P}\left\{(1-\varepsilon)\|\tilde{\mathbf{v}}\|_2^2 \le \left\|\hat{\mathbf{M}}\tilde{\mathbf{v}}\right\|_2^2 \le (1+\varepsilon)\|\tilde{\mathbf{v}}\|_2^2\right\} \ge 1-4e^4\exp(-M\varepsilon/16K) \quad \text{(B.5)}$$

with $\varepsilon \ge 64K/M$.

Now, we generalize the result of (B.5) to any $K$-sparse $\tilde{\mathbf{v}} \in \tilde{\mathbf{V}}_\Lambda$. Define $\mathbf{e}_i = [e_i(1), e_i(2), \cdots, e_i(N)]^{\mathrm{T}} \in \mathbb{R}^N$ with

$$e_i(j) = \begin{cases} 1 & i = j \\ 0 & i \ne j \end{cases} \quad \text{(B.6)}$$

For any $\tilde{\mathbf{v}} \in \tilde{\mathbf{V}}_\Lambda$, $\tilde{\mathbf{v}}$ can be represented as $\tilde{\mathbf{v}} = \sum_{i\in\Lambda}\tilde{v}_i\mathbf{e}_i$ with $\sum_{i\in\Lambda}\tilde{v}_i^2 = 1$. Then

$$\begin{aligned} \left\|\hat{\mathbf{M}}\tilde{\mathbf{v}}\right\|_2^2 &= \left\|\hat{\mathbf{M}}\sum_{i\in\Lambda}\tilde{v}_i\mathbf{e}_i\right\|_2^2 \\ &= \sum_{i\in\Lambda}\tilde{v}_i^2\left\|\hat{\mathbf{M}}\mathbf{e}_i\right\|_2^2 + \sum_{\substack{i,j\in\Lambda \\ i\ne j}}\tilde{v}_i\tilde{v}_j\left\langle\hat{\mathbf{M}}\mathbf{e}_i, \hat{\mathbf{M}}\mathbf{e}_j\right\rangle \end{aligned} \quad \text{(B.7)}$$

Note that

$$\begin{aligned} &\left\langle\hat{\mathbf{M}}\mathbf{e}_i, \hat{\mathbf{M}}\mathbf{e}_j\right\rangle \\ &= \frac{1}{2}\left[\left\|\hat{\mathbf{M}}\frac{\mathbf{e}_i+\mathbf{e}_j}{\sqrt{2}}\right\|_2^2 - \left\|\hat{\mathbf{M}}\frac{\mathbf{e}_i-\mathbf{e}_j}{\sqrt{2}}\right\|_2^2\right] \end{aligned} \quad \text{(B.8)}$$

By applying (B.5), we can establish the CoM inequalities for the vectors $\mathbf{e}_i$, $(\mathbf{e}_i+\mathbf{e}_j)/\sqrt{2}$ and $(\mathbf{e}_i-\mathbf{e}_j)/\sqrt{2}$, i.e.,

$$\begin{aligned} &\mathbb{P}\left\{(1-\varepsilon) \le \left\|\hat{\mathbf{M}}\mathbf{e}_i\right\|_2^2 \le (1+\varepsilon)\right\} \\ &\ge 1-4e^4\exp(-M\varepsilon/16) \qquad (\varepsilon \ge 64/M) \end{aligned} \quad \text{(B.9)}$$

$$\begin{aligned} &\mathbb{P}\left\{(1-\varepsilon) \le \left\|\hat{\mathbf{M}}\frac{\mathbf{e}_i+\mathbf{e}_j}{\sqrt{2}}\right\|_2^2 \le (1+\varepsilon)\right\} \\ &\ge 1-4e^4\exp(-M\varepsilon/32) \qquad (\varepsilon \ge 128/M) \end{aligned} \quad \text{(B.10)}$$

$$\begin{aligned} &\mathbb{P}\left\{(1-\varepsilon) \le \left\|\hat{\mathbf{M}}\frac{\mathbf{e}_i-\mathbf{e}_j}{\sqrt{2}}\right\|_2^2 \le (1+\varepsilon)\right\} \\ &\ge 1-4e^4\exp(-M\varepsilon/32) \qquad (\varepsilon \ge 128/M) \end{aligned} \quad \text{(B.11)}$$

According to (B.9)~(B.11), we define the following events:

$$\mathrm{A} = \left\{1-\varepsilon \le \left\|\hat{\mathbf{M}}\tilde{\mathbf{v}}\right\|_2^2 \le 1+\varepsilon,\ \tilde{\mathbf{v}} \in \tilde{\mathbf{V}}_\Lambda\right\} \quad \text{(B.12)}$$

$$\mathrm{B}_i = \left\{(1-\varepsilon/K) \le \left\|\hat{\mathbf{M}}\mathbf{e}_i\right\|_2^2 \le (1+\varepsilon/K)\right\} \qquad (i \in \Lambda) \quad \text{(B.13)}$$

$$\mathrm{C}_{ij} = \left\{(1-\varepsilon/K) \le \left\|\hat{\mathbf{M}}\frac{\mathbf{e}_i+\mathbf{e}_j}{\sqrt{2}}\right\|_2^2 \le (1+\varepsilon/K)\right\} \quad (i, j \in \Lambda,\ i \ne j) \quad \text{(B.14)}$$

$$\mathrm{D}_{ij} = \left\{(1-\varepsilon/K) \le \left\|\hat{\mathbf{M}}\frac{\mathbf{e}_i-\mathbf{e}_j}{\sqrt{2}}\right\|_2^2 \le (1+\varepsilon/K)\right\} \quad (i, j \in \Lambda,\ i \ne j) \quad \text{(B.15)}$$

with

$$\mathbb{P}(\mathrm{B}_i) = 1-4e^4\exp(-M\varepsilon/16K),$$
$$\mathbb{P}(\mathrm{C}_{i,j}) = 1-4e^4\exp(-M\varepsilon/32K).$$
$$\mathbb{P}(\mathrm{D}_{i,j}) = 1-4e^4\exp(-M\varepsilon/32K).$$

Our aim is to derive the probability $\mathbb{P}(\mathrm{A})$ based on $\mathbb{P}(\mathrm{B}_i)$, $\mathbb{P}(\mathrm{C}_{ij})$ and $\mathbb{P}(\mathrm{D}_{ij})$. Combing (B.13)~(B.15) with (B.7), we can derive the following result:

$$\begin{aligned} \left\|\hat{\mathbf{M}}\tilde{\mathbf{v}}\right\|_2^2 &\le (1+\varepsilon/K)\sum_{i\in\Lambda}\tilde{v}_i^2 + \varepsilon/K\sum_{\substack{i,j\in\Lambda \\ i\ne j}}\left|\tilde{v}_i\tilde{v}_j\right| \\ &= 1+\varepsilon/K\sum_{i,j\in\Lambda}\left|\tilde{v}_i\tilde{v}_j\right| \\ &\le 1+\varepsilon/K\left(\sum_{i\in\Lambda}|\tilde{v}_i|\right)^2 \\ &\le 1+\varepsilon \end{aligned} \quad \text{(B.16)}$$

Similarly, we can also derive that

$$\left\|\hat{\mathbf{M}}\tilde{\mathbf{v}}\right\|_2^2 \ge 1-\varepsilon \quad \text{(B.17)}$$

(B.16) and (B.17) demonstrate that the $1-\varepsilon \le \left\|\hat{\mathbf{M}}\tilde{\mathbf{v}}\right\|_2^2 \le 1+\varepsilon$ holds for any $\tilde{\mathbf{v}} \in \tilde{\mathbf{V}}_\Lambda$ when (B.13)~(B.15) are satisfied. As a result, the probability $\mathbb{P}(\mathrm{A})$ can be derived from $\mathbb{P}(\mathrm{B}_i)$, $\mathbb{P}(\mathrm{C}_{ij})$ and $\mathbb{P}(\mathrm{D}_{ij})$:

$$\begin{aligned} \mathbb{P}(\mathrm{A}) &\ge \mathbb{P}\left(\left(\bigcap_{i\in\Lambda}\mathrm{B}_i\right)\cap\left(\bigcap_{i,j\in\Lambda, i\ne j}\mathrm{C}_{ij}\right)\cap\left(\bigcap_{i,j\in\Lambda, i\ne j}\mathrm{D}_{ij}\right)\right) \\ &= 1-\mathbb{P}\left(\left(\bigcap_{i\in\Lambda}\mathrm{B}_i^c\right)\cap\left(\bigcap_{i,j\in\Lambda, i\ne j}\mathrm{C}_{ij}^c\right)\cap\left(\bigcap_{i,j\in\Lambda, i\ne j}\mathrm{D}_{ij}^c\right)\right) \\ &\ge 1-\sum_{i\in\Lambda}\mathbb{P}(\mathrm{B}_i^c) - \sum_{\substack{i,j\in\Lambda \\ i\ne j}}\mathbb{P}(\mathrm{C}_{ij}^c) - \sum_{\substack{i,j\in\Lambda \\ i\ne j}}\mathbb{P}(\mathrm{D}_{ij}^c) \\ &= 1-K\mathbb{P}(\mathrm{B}_i^c) - K(K-1)\mathbb{P}(\mathrm{C}_{ij}^c) - K(K-1)\mathbb{P}(\mathrm{D}_{ij}^c) \\ &\ge 1-8e^4K^2\exp(-M\varepsilon/32K) \end{aligned} \quad \text{(B.18)}$$

where $\varepsilon \ge 128K/M$. We prove the results of Lemma 2. ■

## REFERENCES

[1] R. G. Vaughan, N. L. Scott, and D. R. White, “The theory of bandpass sampling,” *IEEE Trans. Signal Process.*, vol. 39, no.9, pp. 1973-1984, 1991.

[2] K. C. Ho, Y. T. Chan, and R. Inkol, “A digital quadrature demodulation system,” *IEEE Trans. Aerospace and Electronic Systems*, vol. 32, no. 4, pp. 1218-1227, 1996.

[3] E. Candès, T. Tao, “Decoding by linear programming,” *IEEE Trans. Inf. Theory*, vol. 51, no.12, pp. 4203-4215, 2005.

[4] D. Donoho, “Compressed sensing,” *IEEE Trans. Inf. Theory*, vol.52, pp. 1289-1306, 2006.

[5] E. Candès, J. Romberg, and T. Tao, “Robust uncertainty principles: exact signal reconstruction from highly incomplete frequency information,” *IEEE Trans. Inf. Theory*, vol. 52, no.2, pp. 489-509, 2006.

[6] J. Laska, S. Kirolos, Y. Massoud, R. Baraniuk, A. Gilbert, M. Iwen, and M. Strauss, “Random sampling for analog-to-information conversion of wideband signals,” in *Proc. IEEE Dallas Circuits and Systems Workshop (DCAS)*, Dallas, Texas, 2006, pp.119-122.

[7] J. Tropp, M. Wakin, M. Duarte, D. Baron, and R. Baraniuk, “Random filters for compressive sampling and reconstruction,” in *Proc. IEEE Int. Conf. on Acoustics, Speech, and Signal Processing (ICASSP)*, Toulouse, France, 2006.

[8] J. Laska, S. Kirolos, M. Duarte, T. Ragheb, R. Baraniuk, and Y. Massoud, "Theory and implementation of an analog-to-information converter using random demodulation," in *Proc. IEEE Int. Symp. on Circuits and Systems (ISCAS)*, New Orleans, Louisiana, 2007, pp. 1959-1962.

[9] J. A. Tropp, J. N. Laska, M. F. Duarte, J. K. Romberg, and R. G. Baraniuk, "Beyond Nyquist: efficient sampling of sparse bandlimited signals," *IEEE Trans. Inf. Theory*, vol. 56, no. 1, pp.520-544, 2010.

[10] O. Taheri, S. A. Vorobyov, "Segmented compressed sampling for analog-to-information conversion: method and performance analysis," *IEEE Trans. Signal Process.*, vol. 59, no.2, pp. 554-572, 2011.

[11] M. Mishali, Y. C. Eldar, and A. Elron, "Xampling: signal acquisition and processing in union of subspaces," *IEEE Trans. Signal Process.*, vol.59, no.10, pp. 4719-4734, 2011.

[12] D. Donoho and M. Elad, "Optimally sparse representation in general (nonorthogonal) dictionaries via l1 minimization," *Proc. Natl. Acad. Sci.*, vol.100, no.5, pp.2197-2202, 2003.

[13] G. Shi, J. Lin, X. Chen, F. Qi, D. Liu, and L. Zhang, "UWB echo signal detection with ultra-low rate sampling based on compressed sensing," *IEEE Trans. Circuits and Systems II: Express Briefs*, vol. 55, no. 4, pp. 379-383, 2008.

[14] E. Candès, "The restricted isometry property and its implications for compressed sensing," *Comptes rendus Mathematigue*, vol.346, no.9-10, pp.589-592, 2008.

[15] M. A. Davenport, J. N. Laska, J. R. Treichler, R. G. Baraniuk, "The pros and cons of compressive sensing for wideband signal acquisition: noise folding vs. dynamic range," *IEEE Trans. Signal Process.*, vol.60, no.9, pp.4628-4642, 2012.

[16] E. Arias-Castro and Y. C. Eldar, "Noise folding in compressed sensing," *IEEE Signal Processing Letters*, vol.18, no.8, pp.478-481, 2011.

[17] Y. Chi, L. L. Scharf, A. Pezeshki, and A. R. Calderband, "Sensitivity to basis mismatch in compressed sensing," *IEEE Trans. Signal Process.*, vol. 59, no. 5, pp. 2182-2195, 2011.

[18] J. A. Tropp, A. C. Gilbert, "Signal recovery from partial information via orthogonal matching pursuit," *IEEE Trans. Inf. Theory*, vol. 53, no. 12, pp.4655-4666, 2007.

[19] D. Needell, J. A. Tropp, "CoSaMP: Iterative signal recovery from incomplete and inaccurate samples," *Appl. Comput. Harmon. Anal.*, vol.26, no.3, pp.301-321, 2009.

[20] W. Yin, S. Osher, D. Goldfarb, and J. Darbon, "Bregman iterative algorithm for l1-minimization with applications to compressive sensing," *SIAM J. Imaging Sci.*, vol.1, no.1, pp.143-168, 2008.

[21] S. R. Becker, E. J. Candès, and M. C. Grant, "Templates for convex cone problems with applications to sparse signal recovery," *Mathematical Programming Computation*, vol.3, no.3, pp.165-218, 2011.

[22] M. Fornasier, "Numerical methods for sparse recovery," *Radon Series Comp. Appl. Math*, vol.9, pp.1-110, 2010.

[23] E. van den Berg and M. P. Friedlander, "Probing the Pareto frontier for basis pursuit solutions," *SIAM J. Sci. Comput.*, vol.31, no.2, pp.890-912, 2008.

[24] E. van den Berg and M. P. Friedlander, "Sparse optimization with least-squares constraints," *SIAM J. Optim.*, vol.21, no.4, pp.1201-1229, 2011.

[25] http://www.cs.ubc.ca/~mpf/spgl1/index.html

[26] M. Ledoux, *The Concentration of Measure Phenomenon*. New York: American Mathematical Society, 2001.

[27] S. Dasgupta and A. Gupta, "An elementary proof of a theorem of Johnson and Lindenstrauss," *Random Struct. Algor.*, vol. 22, no. 1, pp. 60-65, 2003.

[28] D. Achlioptas, "Database-friendly random projections," in *Proc. 20th ACM Symp. Principles of Database Systems*, Santa Barbara, CA, 2011, pp.274-281.

[29] R. G. Baraniuk, M. A. Davenport, R. DeVore, and M. B. Wakin, "A simple proof of the restricted isometry property for random matrices," *Construct. Approx.*, vol. 28, no. 3, pp. 253-263, 2008.

[30] J. Y. Park, H. L. Yap, C. J. Rozell, and M. B. Wakin, "Concentration of measure for block diagonal matrices with applications to compressive signal processing," *IEEE Trans. Signal Process.*, vol. 59, no. 12, pp. 5859-5875, 2011.

[31] B. M. Sanandaji, T. L. Vincent, and M. B. Wakin, "Concentration of measure inequalities for Toeplitz matrices with applications," *IEEE Trans. Signal Process.*, vol. 61, no. 1, pp. 109-117, 2013.

[32] Y. Xiang, L. Li, and F. Li, "Compressive sensing by white random convolution," 2009 [Online]. Available: http://arxiv.org/abs/0909.2737.

[33] M. Talagrand, "A New Look at independence," *Annal. Prob.*, vol.24, no.1, pp.1-34, 1996.

[34] M. Davenport, "Random observations on random observations: Sparse signal acquisition and processing," PhD thesis, ECE Department, Rice University, Aug. 2010.

[35] Nadav Levanon, and Eli Mozeson, *Radar signals*. Hoboken, New Jersey : John Wiley & Sons, Inc., 2004, pp. 53-73.

[36] R. Frank, "Polyphase codes with good nonperiodic correlation properties," *IEEE Trans. Inf. Theory*, vol. 9, no. 1, pp. 43-45, 1963.

[37] D. Chu, "Polyphase codes with good periodic correlation properties," *IEEE Trans. Inf. Theory*, vol. 18, no. 4, pp. 531-532, 1972.

[38] O. Bar-Ilan, and Y. C. Eldar, "Sub-Nyquist radar via Doppler focusing," 2012 [Online]. Available: http://arxiv.org/abs/1211.0722.

[39] D. M. Malioutov, M. Cetin, and A. S. Willsky, "Sparse signal reconstruction perspective for source localization with sensor arrays," *IEEE Trans. Signal Process.*, vol.53, no.8, pp.3010-3022, 2005.

[40] C. Ekanadham, D. Tranchina, and E. P. Simoncelli, "Recovery of sparse translation-invariant signals with continuous basis pursuit," *IEEE Trans. Signal Process.*, vol.59, no.10, pp.4735-4744, 2011.

[41] M. A. Davenport, P. T. Boufounos, M. B. Wakin, and R. G. Baraniuk, "Signal processing with compressive measurements," *IEEE J. Selected Topics in Signal Processing*, vol.4, no.2, pp. 445-460, 2010.

[42] P. T. Boufounos, and M. Salman Asif, "Compressive sensing for streaming signals using the streaming greedy pursuit," In *Proc. 2010 Military Communications Conference(MILCOM 2010)*, San Jose, CA, 2010, pp. 1205-1210.

[43] M. Salman Asif, and J. Romberg, "Sparse recovery of streaming signals using L1-homotopy," 2013 [Online]. Available: http://arxiv.org/abs /1306.3331.